\documentclass[sigplan]{acmart}

\usepackage{wrapfig}
\usepackage{comment}
\usepackage{colortbl}

\usepackage{hyperref}

\usepackage{color,soul}
\usepackage{graphics}
\usepackage{hyperref}
\usepackage{lipsum}
\usepackage{tikz}
\usepackage{enumitem}	% for fine control of list env
\usepackage{listings} % for code listing
\usepackage{booktabs}	% xzl: for fancy tables
\usepackage{lipsum}

\usepackage{capt-of}	% xzl: used to override the ``figure'' and ``table'' in caption.

\usepackage{todonotes}  %xzl: for "missing figures"
\usepackage{subcaption} %xzl: for subfigure. incompatible with package subfig
\usepackage{mathrsfs}	% xzl: for Ralph Smith’s Formal Script Font (rsfs)
\usepackage{amsfonts}
\usepackage{inconsolata} % xzl: this uses inconsolata for texttt

\usepackage[export]{adjustbox} % xzl: flush figure to left/right

\usepackage{multirow} % xzl: for rotate text in table

\usepackage[normalem]{ulem} % hym: for strikethrough

\newcommand*\circlew[1]{\tikz[baseline=(char.base)]{
		\node[shape=circle,fill=white,draw,text=black,inner sep=1pt] (char) {#1};}}
				
  {\begin{list}{$\bullet$}%
     {\setlength{\parsep}{0pt}%
      \setlength{\topsep}{0pt}%
      \setlength{\itemsep}{2pt}}}%
  {\end{list}}
\newcommand\note[1]{\sethlcolor{yellow} \hl{#1}} % highlighted notes of other colors.
\newcommand\noted[1]{} % remove highlights.

\newcommand\ksm[1]{{\color{blue}{ksm: #1}}}
\newcommand\sect[1]{Section~\ref{sec:#1}}

\newenvironment{myitemize}%
  {\begin{itemize}
	[leftmargin=0cm,
		itemindent=.3cm,
		labelwidth=\itemindent,
		labelsep=0pt,
		parsep=3pt,
		topsep=2pt,
		itemsep=1pt,
		align=left]
  }%
  {\end{itemize}}

\newenvironment{myenumerate}%
  {\begin{enumerate}
	[leftmargin=0cm,itemindent=.5cm,labelwidth=\itemindent,
		labelsep=0pt,
		parsep=1pt,
		topsep=1pt,
		itemsep=3pt,
		align=left]
  }%
  {\end{enumerate}}

\newcommand{\code}[1]{\texttt{\small{#1}}}	
\newcommand{\sys}{StreamBox-HBM} % Streambox

\newcommand{\spe}{stream analytics engine}
\newcommand{\hbma}{KPA}

\newcommand{\reorg}{grouping}

\newcommand{\kv}{$\langle$key,value$\rangle$}

\newcommand{\sort}{$Sort$}
\newcommand{\hash}{$Hash$}

\makeatletter
\def\@copyrightspace{\relax}
\makeatother

\usepackage{epsfig,endnotes}
\usepackage{url}
\usepackage{multirow}
\usepackage{balance} % For balanced columns on the last page
\usepackage{etoolbox}
\usepackage{booktabs} % For formal tables

\newcommand*{\affaddr}[1]{#1} % No op here. Customize it for different styles.
\newcommand*{\affmark}[1][*]{\textsuperscript{#1}}

\copyrightyear{2019}
\acmYear{2019}
\setcopyright{acmcopyright}
\acmConference[ASPLOS '19]{2019 Architectural Support for Programming Languages
and Operating Systems}{April 13--17, 2019}{Providence, RI, USA}
\acmBooktitle{2019 Architectural Support for Programming Languages and Operating
Systems (ASPLOS '19), April 13--17, 2019, Providence, RI, USA}
\acmPrice{15.00}
\acmDOI{10.1145/3297858.3304031}
\acmISBN{978-1-4503-6240-5/19/04}

\begin{document}

%don't want date printed
\date{}
\sloppy

\title{StreamBox-HBM: Stream Analytics on High Bandwidth Hybrid Memory}
\renewcommand{\shorttitle}{StreamBox-HBM}

\iffalse
\author{
\rm Hongyu Miao\affmark[1]~~~  
\rm Myeongjae Jeon\affmark[2]~~~
\rm Gennady Pekhimenko\affmark[3]~~~ 
\rm Kathryn S. McKinley\affmark[4]~~~ 
\rm Felix Xiaozhu Lin\affmark[1]\\
\affaddr{\affmark[1]Purdue ECE~~~~~~~~
          \affmark[2]UNIST ~~~~~~~~~~
          \affmark[3]University of Toronto ~~~~~~~~~~
          \affmark[4]Google}
%\vspace{-2ex}
}
\fi

%\iffalse
\author{Hongyu Miao}
\affiliation{
	\institution{Purdue ECE}
}
%\email{miaoh@purdue.edu}

\author{Myeongjae Jeon}
	\affiliation{
	  \institution{UNIST}
}
%\email{mjjeon@unist.ac.kr}

\author{Gennady Pekhimenko}
\affiliation{
	\institution{University of Toronto}
}
%\email{miaoh@purdue.edu}

\author{Kathryn S. McKinley}
\affiliation{
	\institution{Google}
}
%\email{miaoh@purdue.edu}

\author{Felix Xiaozhu Lin}
\affiliation{
	\institution{Purdue ECE}
}
%\email{miaoh@purdue.edu}
%\fi

\renewcommand{\shortauthors}{H. Miao, M. Jeon, G. Pekhimenko, K. S. McKinley, F. X. Lin }

% !TeX root = main.tex

%\subsection*{Abstract}
\begin{abstract}
Stream analytics has an insatiable demand for memory and performance.
%Unfortunately, current stream processing systems still struggle to exploit
Emerging \emph{hybrid} memories combine commodity DDR4 DRAM 
with 3D-stacked High Bandwidth Memory (HBM) DRAM to meet such
demands.  % This
% paper is seeking to answer the question of "Can stream analytics
% benefit from HBM?", and the answer is "Yes". 
However, achieving this promise is challenging because (1) HBM is
capacity-limited and (2) HBM boosts performance best for sequential
access and high parallelism workloads.  At first glance, stream
analytics appears a particularly poor match for HBM because they have
high capacity demands and data grouping operations, their most
demanding computations, use random access.

This paper presents the design and implementation of StreamBox-HBM, a stream
analytics engine that exploits hybrid memories to achieve scalable
high
performance. % To address these challenges, we present \sys{}, a stream analytics engine optimized for hybrid memories
%Contributions of this system include (1) 
StreamBox-HBM performs data grouping with
sequential access sorting algorithms in HBM, in contrast to random
access hashing algorithms commonly used in DRAM. %existing stream engines,
%(2) by \sout{compressing} \hym{extracting} stream records in HBM to keys and record pointers in a \emph{Key Pointer Array} (KPA) data structure.
%(2) 
StreamBox-HBM solely uses HBM to store \emph{Key Pointer Array} (KPA) data
structures that contain only partial records (keys and pointers to full
records) for grouping operations.
% StreamBox-HBM  
It dynamically creates and manages prodigious data and pipeline
parallelism, choosing when to allocate KPAs in HBM. It dynamically optimizes for
both the high bandwidth and limited capacity of HBM, and the limited
bandwidth and high
capacity of standard DRAM.
%Our contributions are: (1) the idea for HBM's limited capacity that HBM will be solely used to keep a smaller \emph{Key Pointer Array} (KPA) data structure, which includes extracted keys and pointers to complete records; (2) the optimization for HBM's high bandwidth that sequential algorithms (e.g. sort) are used to implement grouping operators; (3) a novel runtime that dynamically manages pipeline parallelism and KPA placement based on both HBM's high bandwidth and limited capacity and DRAM's high capacity and limited bandwidth.
%On a Intel Knights Landing machine,
% \sys{} achieves 80 million records per second throughput and 238 GB/s bandwidth utilization, which
%is within 59\% of peak bandwidth.

StreamBox-HBM achieves 110 million records per second % throughput 
and 238 GB/s memory bandwidth while effectively utilizing all 64 
 cores of Intel's Knights Landing, a commercial server with hybrid memory.
% in a scalable manner.
It outperforms
stream engines with sequential access algorithms without KPAs by 7$\times$ and stream engines with random access algorithms by an order of magnitude in throughput. 
%Comparing with distributed engines, \sys{} can achieve an order of magnitude of throughput\note{updated}. 
To the best of our knowledge, StreamBox-HBM is the first stream engine optimized for hybrid memories. %and achieves the best performance on a single node.
\end{abstract}



\maketitle

% !TeX root = main.tex

%%%%%%%%%%%%%%%%%%%%%%%%%%%%%%%%%%%%%%%%%%%%%%%
%
% xzl: This is the Intro. It can be produced as part of the paper PDF and a standalone html.
% -- Don't put any non standard latex commands here. No need to save space.
%
% html: see summary/
%%%%%%%%%%%%%%%%%%%%%%%%%%%%%%%%%%%%%%%%%%%%%%%

\newcommand{\isolation}{Enforcing isolation with ARM TrustZone}
\newcommand{\noconcurrency}{Exposing low-level abstractions of trusted computations}
\newcommand{\seqmm}{Unbounded arrays as the universal memory abstraction}
\newcommand{\hints}{Exploiting performance hints from the untrusted}

\section{Introduction}
\label{sec:intro}

%\ksm{Three todo items: (i) Change all figure references to be consistent. We currently have 7(a) and 7a.  I prefer 7(a) because it is consistent with the labeling in the figure itself. (ii) Order the citations (10, 28, 45) versus random (10, 45, 28). It makes it easier for the reader to look them up. (iii) Add a url citation for the open source version of StreamBox and StreamBox-H (if they are different add two).}

% --- stream analytics bkgnd ---- %

% Modern data analytics % requires high throughput stream processing.
% use
Cloud analytics and the rise of the Internet of Things 
increasingly challenge stream analytics engines to achieve high throughput (tens of million records per
second) and low output delay (sub-second)~\cite{trill,streambox,drizzle,sparkstreaming}.  Modern
engines ingest unbounded numbers of time-stamped data records, continuously push them
through a pipeline of operators, and produce a series of results over
\textit{temporal windows} of records.
%\st{Although record arrival is often bursty,
%each operator must execute continuously, performing computations that are scoped by record time
%in a window.} % and may maintain internal state.
%classic DBMS, where data is standing and queries are flowing.
Many streaming pipelines group data in multiple rounds
(e.g., based on record time and keys) and consume grouped data with a single-pass reduction (e.g., computing average values per key).
For instance, data center analytics compute the distribution of machine
utilization and network request arrival rate, and then join them by time.
Data \reorg{} often consumes a majority of the execution time and is
crucial to low output delay in production systems such as Google
Dataflow ~\cite{google-dataflow} and Microsoft
Trill~\cite{trill}. Grouping operations dominate queries in TPC-H (18 of 22) ~\cite{tpc-h}, BigDataBench (10 of 19)~\cite{bigdatabench}, AMPLab Big Data Benchmark (3 of 4)~\cite{amplab-benchmark}, and even Malware Detection~\cite{malware_detection}.
%\sout{These challenges require stream engines to be carefully engineered for the concurrency and memory systems of modern hardware.} 
These challenges require stream engines to carefully choose algorithms (e.g. Sort vs. Hash) and data structures for data grouping to harness the concurrency and memory systems of modern hardware.
% \st{to deliver results continuously by} effective exploitation \note{of modern hardware features}. 
% \st{concurrency and hardware features of modern memory systems.} 
% , even as data arrival is
% bursty and is much more challenging than traditional batch analytics.
% %In executing a pipeline, the engine spends most of its time on moving data in memory.
% \note{only use record time; no event time}

%We seek to accelerate stream analytics with emerging high-bandwidth
%memory (HBM).

Emerging 3D-stacked memories, such as high-bandwidth memory (HBM), offer opportunities and
challenges for modern workloads and stream analytics. % in particular.
%For example, 
HBM delivers much %an order of magnitude 
higher bandwidth (several hundred
GB/s) than DRAM, but at longer latencies and 
at reduced % an order of magnitude
capacity (16~GB) versus hundreds of GBs of DRAM. %DRAM TBs. % \st{We show exercising this high
% bandwidth requires prodigious hardware and software
% parallelism, and also software with sequential access and data reuse.} \mj{redundant to the last sentence of the next paragraph (and partially with prev paragraph. I'd rather delete this sentence.}
Modern CPUs (KNL~\cite{intel-knl}), GPUs (NVIDIA Titan V~\cite{nvidia_titanv}), FPGAs (Xilinx Virtex UltraScale+~\cite{xilinx_virtex}), and Cloud TPUs (v2 and v3~\cite{cloud_tpu}) are using HBM/HBM2.
Because of HBM capacity limitations, vendors couple HBM and
standard DRAM in hybrid memories on platforms such as 
%IBM Power 7 and 8~\cite{ibm-power8}, Intel Haswell~\cite{intel-haswell}, Broadwell, Skylake,  %hym: these citations are wrong...
%\note{GPUs/FPGAs are using HBM, but not hybrid memory}
Intel Knights Landing~\cite{intel-knl}.  
%These systems provision rich hardware parallelism at all levels. 
%HBM's slightly longer latency poses
%a different optimization problem compared to the traditional faster
%tiers in other memory systems (e.g., SRAM or local NUMA nodes) that
%offer both higher bandwidth and \textit{lower} latency.
Although researchers have achieved substantial improvements for high
performance computing~\cite{li17sc,7965110} and machine
learning~\cite{you17sc} on hybrid HBM and DRAM 
systems, optimizing streaming for hybrid memories is more challenging.
%Streaming requires substantial networking for ingress and throughput. % and high % capacity.
Streaming queries require high network bandwidth for ingress and high throughput for the whole pipeline.  
Streaming computations are dominated by data grouping, which
currently use hash-based data structures and random access algorithms.
We demonstrate these challenges with measurements on Intel's Knights
Landing  architecture (\S\ref{sec:bkgnd}). Delivering high throughput
and low latency streaming on 
HBM requires high degrees of
software and hardware parallelism and
sequential accesses.
 
We present \sys{}, a  stream analytics engine that transforms streaming data
and computations to exploit hybrid
HBM and DRAM memory systems. It performs sequential data grouping computations primarily in HBM.
%These operations typically dominate the computational needs of stream analytics.
% To reduce the capacity needs of data in HBM, 
%\sout{\sys{} dynamically compresses records by allocating required keys and pointers to complete records on HBM in a data structure we call Key Pointer Arrays (KPAs).}
\sys{} dynamically extracts into HBM one set of keys at a
time together with pointers to complete records in a data structure we
call \emph{Key Pointer Array} (KPA), minimizing the use of precious
HBM memory capacity.
To produce sequential accesses, we implement grouping computations as
sequential-access parallel sort, merge, and join with wide vector instructions % (\code{avx512})
on KPAs in a streaming algorithm library.
These algorithms are best for HBM and differ from hash-based grouping on DRAM in other engines~\cite{beam,dataflowAkidau2015,flink,sparkstreaming,streambox}. %gennady: let's cite most common hash-based grouping

\sys{} dynamically manages applications' streaming pipelines.
At ingress, \sys{} allocates records in DRAM.
For grouping computations for key $k$, it dynamically allocates 
extracted KPA records for $k$
on HBM. For other streaming computations such as reduction, \sys{}
allocates and operates on \emph{bundles} of complete records stored in DRAM.
Based on windows of records specified by the pipeline, the \sys{}
runtime further divides each window into bundles to expose data
parallelism in bottleneck stream operations. It uses bundles as the
unit of computation, assigning records to bundles and threads to bundles
or KPA. It detects bottlenecks and dynamically uses out-of-order data and
pipeline parallelism  to optimize throughput and latency by producing sufficient
software parallelism to match hardware capabilities.

The \sys{} runtime monitors HBM
capacity and DRAM bandwidth (the two major resource constraints of hybrid memory)
and optimizes their use to improve performance.
% and protect against the
%limiting factors in hybrid memories -- HBM capacity and DRAM
%bandwidth.
It prevents either resource from becoming a bottleneck with a
single control knob: a decision on where to allocate new KPAs.
By default \sys{} allocates \hbma{}s on HBM.
When the HBM capacity runs low, \sys{} gradually increases the
fraction of new \hbma{}s it allocates on DRAM, adding pressure to the DRAM bandwidth but without saturating it.

We evaluate \sys{} on a 64-core Intel Knights
Landing with 3D-stacked HBM and DDR4 DRAM~\cite{jeffers2016intel} and
a 40 Gb/s Infiniband with RDMA for data ingress.
On 10 benchmarks, \sys{}
achieves throughput up to 110 M records/s (2.6 GB/s)
with an output delay under 1 second. 
%Compared to popular engines, this throughput is XXX\note{FIXME} higher.
%This throughput is on-par with distributed engines on medium-size clusters.
We compare \sys{} to Flink~\cite{flink} on the popular %\st{YCSB} \
YSB
benchmark~\cite{ycsb} % on the same Knights Landing machine, 
where \sys{} achieves 18$\times$ higher throughput per core. 
%\sout{Using RDMA ingestion with \sys{}, that Flink does not support,
%\sys{} achieves 4$\times$ higher total throughput than Flink.} 
Much prior work reports results without data ingress~\cite{trill,streambox}.
As far as we know, \sys{} achieves the best reported records per
second for streaming \emph{with ingress} on a single machine. 
%We compare \sys{} on the same hardware (KNL) and YCSB benchmark with Flink~\cite{flink}, and the result shows that \sys{} outperforms Flink by 4.1x throughput although \sys{} is limited by RDMA ingestion. \sys{}'s throughput will be even higher with faster RDMA network, but Flink cannot saturate 10Gb Ethernet yet\note{updated}. 

The key contributions are as follows.
(1) New empirical results find on real hardware that sequential sorting algorithms for grouping  are
best for HBM, in contrast to DRAM, where random
hashing algorithms are best~\cite{sort-vs-hash-2013, sort-vs-hash-2009, polychroniou15sigmod}. Based on this finding, we optimize grouping computations with sequential algorithms. % to benefit from HBM's high bandwidth; 
(2) A dynamic optimization for limited HBM capacity that
%reduces records size in HBM to KPA keys and pointers to complete records.
reduces records to keys and pointers residing in HBM.  
Although key/value separation is not
new~\cite{monetdb,trill,sqlserver-column,lehman1986query,alphasort,db2-column,c-store},
mostly it occurs statically ahead of time, instead of selectively and
dynamically.
 (3) Our novel runtime manages parallelism and KPA placement %on 
%\sout{HBM and DRAM} \hym{
based on both HBM\text{'s} high bandwidth and limited capacity,
and DRAM\text{'s} high capacity and limited bandwidth. 
%{\hym{\sout{No other system dynamically chooses between a reduced data structure for HBM and the full data structure for DRAM, managing computations and data layout for hybrid systems.}}
% It optimizes
% the use of HBM-DRAM hybrid memory by efficiently allocating
% new KPAs at it processes stream data.
  % it is really both, because to "improve" it
                         % exploits the strength of each, HBM
                         % bandwidth and DRAM capacity but it also
                         % manages the weakness: \note{(HBM capacity and DRAM bandwidth?)}
%by where it allocates new KPAs as
%it processes stream data.
The resulting engine achieves high
throughput, scalability, and bandwidth on hybrid memories.
% \note{either delete or make stronger. ksm: I disagree! I think we  believe it will generalize, but it takes more research to know for sure }
Beyond stream analytics, \sys{}'s techniques should 
improve a range of data processing systems, e.g., batch analytics and
key-value stores, on HBM and near-memory architectures~\cite{mondrian}.
% \st{as well as near-memory architectures}  
To our knowledge, \sys{} is the first stream engine for
hybrid memory systems. % \note{check}  
%\ksm{FIXME with a reference to the open source repository: Upon publication, we will open-source \sys{}.} 
The full source code of StreamBox-HBM is available at
%\url{http://xsel.rocks/p/streambox-hbm}.
%hym: same project, use same website
\url{http://xsel.rocks/p/streambox}.

\section{Background \& Motivation}
\label{sec:bkgnd}

This section presents background on our
 stream analytics programming model, runtime, and High Bandwidth
Memory (HBM). Motivating results explore \code{GroupBy}
implementations with sorting and hashing on
HBM. We find merge-sort exploits
HBM's high memory bandwidth with sequential access patterns and
high parallelism, achieving much higher throughput and scalability than hashing on HBM. % or DRAM.

\subsection{Modern Stream Analytics}

\paragraph{Programming model}
We adopt the popular Apache Beam programming model~\cite{beam} used
by stream engines such as Flink~\cite{flink},
Spark Streaming~\cite{sparkstreaming}, and Google
Cloud Dataflow~\cite{dataflowAkidau2015}. These engines all use declarative
stream operators that group, reduce, or do both on stream data such as
those in
Table~\ref{tab:operators}. % presents some of the common operators. 
To define a stream pipeline, programmers declaratively specify
operators (computations) and a pipeline of how data flows between operators, as
shown in the following pseudo code.

\lstset{
	%language=C++,
	%basicstyle=\ttfamily\footnotesize\small,
	basicstyle=\fontsize{9}{9}\selectfont\ttfamily,
	xleftmargin=1.0ex,
	framexleftmargin=1.0ex,
	frame=tb, % top & bottom
	breaklines=true,
	captionpos=b,
%	numbers=left,
	label=list:correct-spi,
	%belowskip=-1.0 \baselineskip
	}
\begin{lstlisting}[caption= Example Stream Program. It sums up values for each key in every 1-second fixed-size window.]
/* 1. Declare operators */
Source source(/* config info */);
WinGroupbyKey<key_pos> wingbk(1_SECOND);
SumPerKey<key_pos,v_pos> sum;
Sink sink;
/* 2. Create a pipeline */
Pipeline p; p.apply(source); 
/* 3. Connect operators */
connect_ops(source, wingbk);
connect_ops(wingbk, sum);
connect_ops(sum, sink);
/* 4. Execute the pipeline */
Runner r( /* config info */ ); 
r.run(p);  
\end{lstlisting}

%\begin{lstlisting}[caption= Example Stream Program \label{listing:grep}]
%// 1. Declare operators
%Source<type> source(/*config info*/);
%WinGropbyKey<type> wingbk(winsize, kpos);
%SumVperKey<type> sumv(kpos, vpos);
%Sink<type> sink();
%
%// 2. Create a pipeline
%Pipeline* p = Pipeline::create(); 
%p->apply(source); //set source
%
%// 3. Connect transforms together
%connect_transform(source, wingbk);
%connect_transform(wingbk, sumv);
%connect_transform(sumv, sink);
%
%// 4. Evaluate the pipeline
%Evaluator eval(/*config info*/);
%eval.run(p);  // run the pipeline
%\end{lstlisting}

% !TeX root = main.tex

\begin{figure}[t!]
	\centering

	\begin{subfigure}[b]{0.45\textwidth}
		\includegraphics[width=1\linewidth]{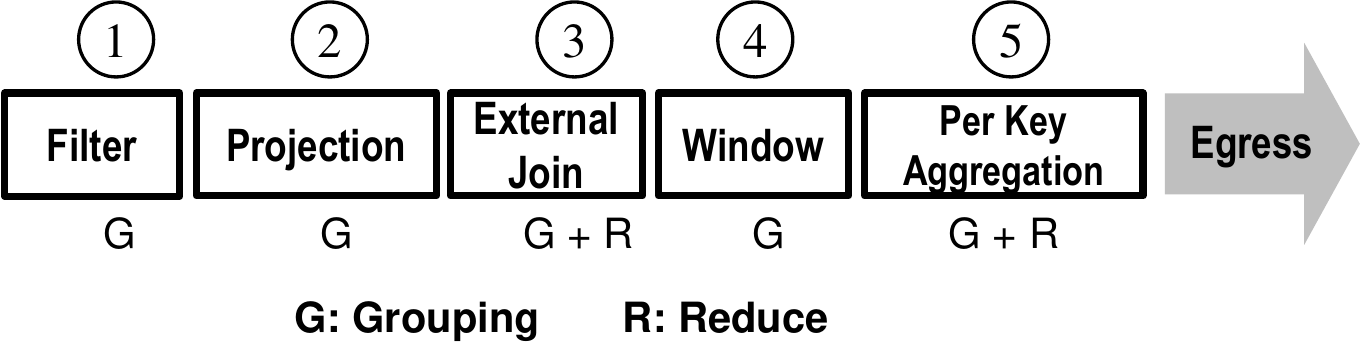}
		\caption{Pipeline of Yahoo streaming benchmark (YSB)~\cite{ycsb} which counts ad views. 
		   It filters records by \textit{ad\_id} \circlew{1}, 
		   takes a projection on columns \circlew{2}, joins by \textit{ad\_id} 
		   with associated \textit{campaign\_id} \circlew{3}, then counts events per campaign per window \circlew{4} \& \circlew{5}. 
		   The pipeline will serve as our running example for design and evaluation (\S\ref{sec:kpa} and \S\ref{sec:eval}).
		}
		\vspace{2mm}
		\label{fig:pipeline-stacked:pipeline} 
	\end{subfigure}

	\begin{subfigure}[b]{0.45\textwidth}
		\includegraphics[width=1\linewidth]{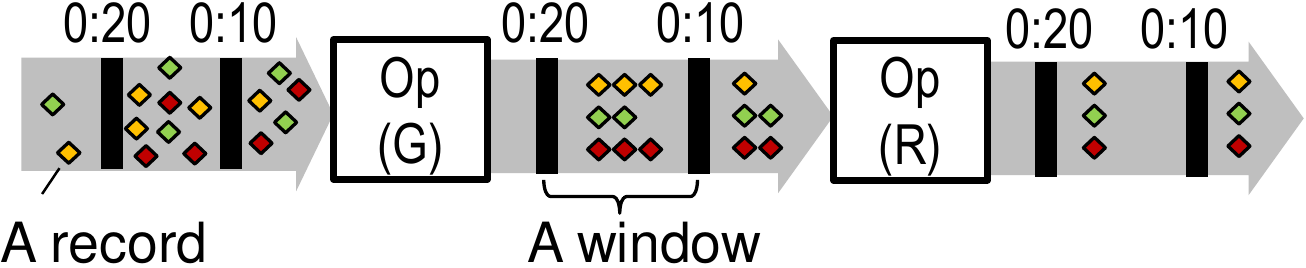}
		\caption{A stream of records flowing through a grouping operator (G) and a reduction operator (R)}
		\vspace{2mm}
		\label{fig:pipeline-stacked:stream} 
	\end{subfigure}
	
	\begin{subfigure}[b]{0.45\textwidth}
		\includegraphics[width=1\linewidth]{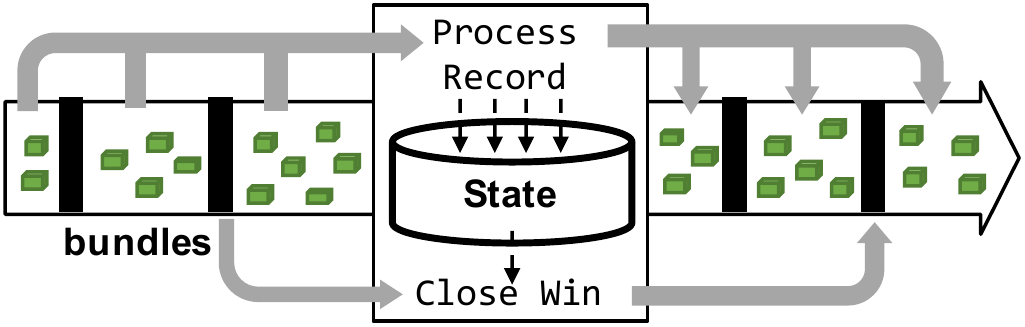}
		\caption{Parallel operator execution. Engine batches records in bundles, consuming and producing bundles in multiple windows in parallel
		}
		\label{fig:pipeline-stacked:op} 
	\end{subfigure}

	\caption{Example streaming data and computations}
	%\vspace{-5pt}		% use as needed
	\label{fig:pipeline-stacked}
\end{figure}
% !TeX root = main.tex

\begin{table}[t!]
\centering
\includegraphics[width=0.45\textwidth]{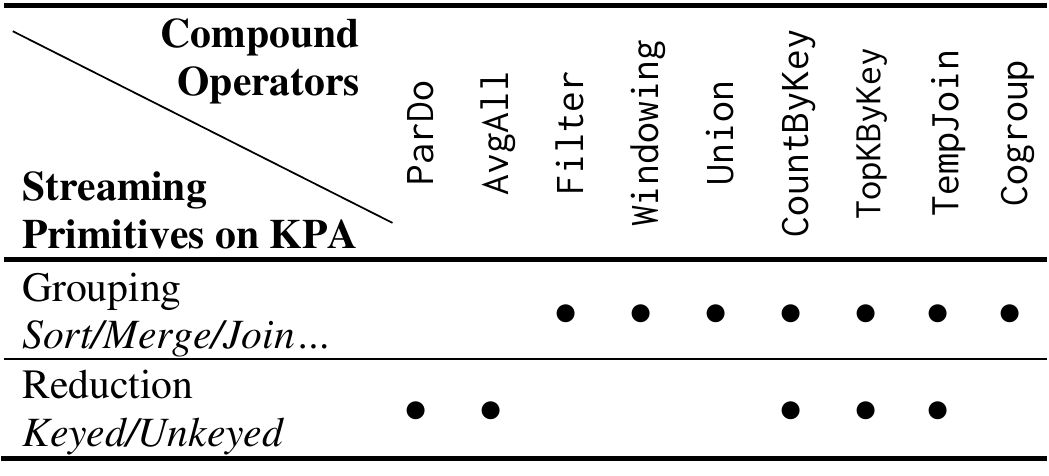}
\caption{Selected compound (declarative) operators in \sys{} and their constituent streaming primitives. 
}
%\vspace{-2pt}		% use as needed
\label{tab:operators}
\end{table}

%\begin{table}[]
%\centering
%
%\begin{tabular}{@{}ll@{}}
%%\hline
%\textsf{\textbf{}}          & \textsf{\textbf{Declarative Operators}} \\ \midrule
%Grouping  & GroupBy, Windowing, CoGroup \\ 
%Reduction & Filter, FlatMap,            \\ 
%(G+R)    & Percentile, TopK, Join      \\ 
%\end{tabular}
%\caption{Common declarative streaming operators \textbf{Hongyu complete}}
%\label{tab:operators}
%\end{table}

\paragraph{Streaming computations: grouping \& reduction}
The declarative operators in Table~\ref{tab:operators} serve two purposes.
(1)  \emph{Grouping} computations organize records by keys and
timestamps contained in sets of records. They sort,
merge, or select a subset of records.
Grouping may both move and compute on records, e.g., by comparing keys.
(2) \emph{Reduction} computations aggregate or summarize existing records
and produce new ones, e.g., by averaging or computing distributions of
values.
% Table~\ref{tab:operators} lists common declarative operators and their constituent computations.
Pipelines may interleave multiple instances of operations, as exemplified in Figure~\ref{fig:pipeline-stacked:pipeline}.
In most pipelines, grouping dominates total execution time.
%\mj{the table does not match with Figure 1b}
%Records inside a pipeline are short-lived, as they are quickly consumed by subsequent operators or externalized to users.
%Grouping operations typically dominate total execution time since records after the reduction are short-lived or immediately externalized to users.
%When the pipeline aggregates
%records, many records and intermediate results are short-lived.
%Grouping operations thus typically dominate total execution time.
%Programmers often specify such computations as a pipeline of declarative \textit{operators}.

\paragraph{Stream execution model}
Figure~\ref{fig:pipeline-stacked:stream} shows our execution model.
Each stream is an unbounded sequence of records $\mathcal{R}$ produced by sources,
such as sensors, machines, or humans.
Each record consists of an event timestamp and an arbitrary number of
attribute keys (\emph{columns}).
%\mj{edited: I don't think watermark is very relevant to our paper's context. Also, it looks the original statement for timestamp is wrong and opposite?}
%Data sources inject \emph{watermarks} into record streams to indicate temporal progress. Upon receiving a watermark, no records with timestamps prior to its timestamp can be pushed into the query pipeline to be processed. However, records may arrive early and can be processed out-of-order~\cite{OutLi2008}.
Data sources inject into record streams special \emph{watermark} records that guarantee
all subsequent record timestamps will be later than the watermark timestamp.
However, records may arrive out-of-order~\cite{OutLi2008}.
%Thus, no records with timestamps later
%than the watermark arrive after it, but records may arrive early and
%thus out-of-order~\cite{OutLi2008}.
A pipeline of stream operations
%\st{$\cal{D}$ = $\{d_1, d_2, \cdots, d_n\}$}
consumes one or more data streams and generates output on temporal windows.
% As records arrive, a \spe{} process them with a pipeline of declarative \textit{operators}.

% xzl: first data mgmt, then thread mgmt
\paragraph{Stream analytics engine}
Stream analytics engines are user-level runtimes that exploit parallelism.
%It pushes data through the pipelines by creating operator tasks, managing
%data and threads, and dynamically mapping them to hardware core and memory resources.
%To amortize scheduling overhead, engines process multiple records at once, e.g., a set of records belonging to the same window.
%To amortize scheduling overhead, engines process a set of records belonging to the same window (a record \textit{bundle}) at once.
They exploit \emph{pipeline parallelism} by executing
multiple operators on distinct windows of records.
We extend the StreamBox engine, which also exploits \emph{data parallelism} by dividing windows into record bundles~\cite{streambox}.
%We build based on the StreamBox runtime~\cite{streambox}, which may further process epochs of windows out-of-order~\cite{OutLi2008} to create epoch parallelism.
%\st{The runtime executes operators that consume and produce bundles of records.}
Figure~\ref{fig:pipeline-stacked:op} illustrates the execution of an
operator. Multiple bundles in multiple windows are processed in parallel.
After finishing processing one window, the runtime closes the window by
combining results from the execution on each bundle in the window.
%In this model, the stream engine exploits both pipeline and data parallelism.

%\mj{shorten from the original}
%Worker threads are shared by operator execution tasks that dispatch disjoint bundles to process.
%The runtime mitigates backpressure by promptly balancing threads among individual operators, achieving good load balancing.

%\begin{comment}
To process bundles, the runtime creates operator tasks, manages threads and data, and maps them to cores and memory resources.
The runtime dynamically varies the parallelism of individual operators
depending on their workloads.
%It further dynamically monitors and controls their memory system behavior by where it allocates KPAs. \note{del?}
At one given moment, distinct worker threads may execute different operators, or execute the same operator on different records.
%\end{comment}

%\mj{we need only two figures in Figure 1. I think Figure 1a can be merged into Figure 1c.}

%automatically adds parallelism to bottleneck operators.

%since statically partitioning hardware resources among operators often leads to load imbalance due to input variation and key skewness.

\subsection{Exploiting HBM}
\label{sec:bkgnd:exploit}

% \note{should introduce knl hw somewhere -- perhaps use a table}. ref~\cite{cheng17cikm} tab 1

%Modern HBM~\cite{hbm,hbm2}, including embedded DRAM~\cite{6816046}, multi-channel DRAM~\cite{intel-knl}, and 3D-stacked DRAM~\cite{die-stacking}, stacks up to 8 DRAM dies on top of one CPU chip.
Modern HBM stacks up to 8 DRAM dies in special purpose silicon chips~\cite{hbm,hbm2}.
%called interposers~\cite{hbm,hbm2}.
Compared to normal DRAM, HBM offers (1) 5--10$\times$ higher bandwidth,
%due to a total wider bus (2014 bits vs. 384 bits for DRAM on KNL);
(2) 5--10$\times$ smaller capacity due to cost and
power~\cite{hbm,hbm2,6757501}, and
(3) latencies typically $\sim$20\% higher due to added stacking silicon layers.

Recent platforms couple HBM and DDR4-based DRAM as a hybrid memory
system~\cite{ibm-power8,intel-haswell,intel-knl}.  Hybrid memories with HBM
and DRAM differ substantially from hybrid memories with SRAM and DRAM; or DRAM
and NVM; or NUMA. In the latter systems, the faster tiers (e.g., on-chip
cache or local NUMA memory) offer both higher bandwidth and \textit{lower}
latency.  HBM lacks a latency benefit. We show next that for workloads to benefit
from HBM, they must exhibit prodigious parallelism and sequential memory access
\textit{simultaneously}. %, as we show in the following experiment.
%HBM differs because it lacks the latency benefit. Furthermore,
%we show below it only benefits workloads that simultaneously exhibit
%prodigious software parallelism and sequential memory
%access.

We measure two versions of GroupBy, a common stream operator on Intel's KNL with
96$\;$GB of commodity DRAM %TODO: add the type of DRAM here: DDR3 or DDR4
 and 16$\;$GB of HBM (Table~\ref{tab:plat}).
(1) \hash{} partitions input \kv{} records and inserts them into an open-addressing, pre-allocated hash table.
(2) \sort{} merge-sorts the input records by key (\S~\ref{sec:kpa:primitives}).
%Each implementation groups 100M \kv{} pairs by key. Keys and values are 64-bit random integers and each key has around 100 values.
We tune both implementations with hardware-specific optimizations and
handwritten vector instructions. % (AVX-512)~\cite{cheng17cikm}.
We derive our Hash from a state-of-the-art
implementation hand-optimized for KNL~\cite{sort-vs-hash-2009}, and
implement Sort from a fast implementation~\cite{cheng17cikm} and hand-optimize it with AVX-512. 
Our \hash{} is 4$\times$ faster (not shown) than a popular, fast hash
table \textit{not} optimized for KNL~\cite{folly}. Both implementations achieve state-of-the-art performance on KNL.

%We run both benchmarks on Intel's Knights Landing which includes 96GB of standard DRAM with peak bandwidth of 75GB/s and 16GB of HBM with an advertised peak bandwidth of 375GB/s. \note{verify}
% !TeX root = main.tex

\begin{figure}[t!]
\centering
   \begin{subfigure}[b]{0.23\textwidth}
   \includegraphics[width=1\linewidth]{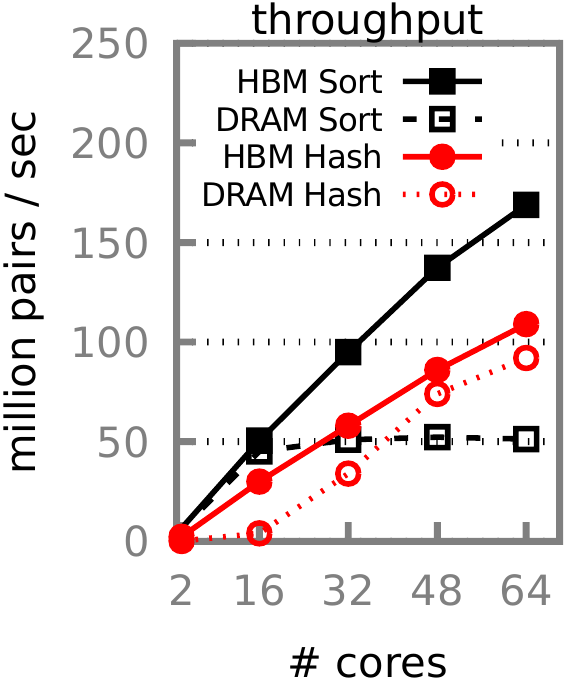}
   % \caption{TBD}
   % Benchmark: cpp drawing tutorial 1
%	\vspace{2mm}
	\end{subfigure}
	\begin{subfigure}[b]{0.23\textwidth}
	   \includegraphics[width=1\linewidth]{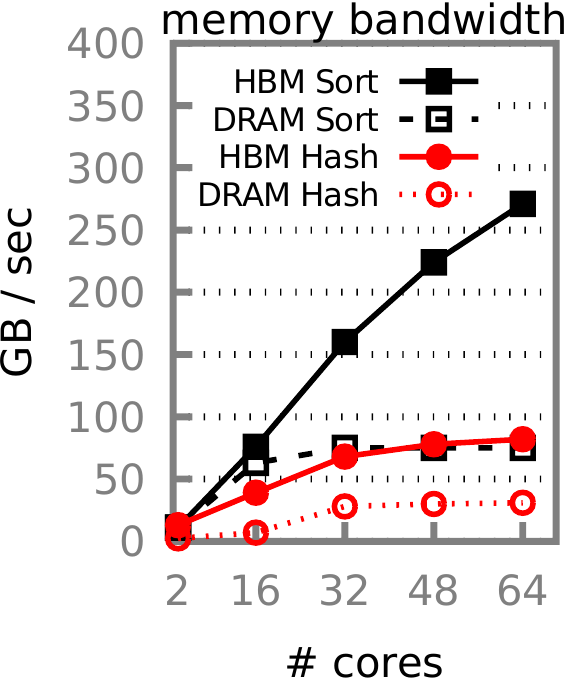}
%	   \caption{TBD}
	\end{subfigure}

        \caption{GroupBy on HBM and DRAM operating on
 100M key/value records with about 100 values per key. Keys and
        values are 64-bit random integers.  
        	Sort leverages HBM bandwidth with sequential access
                and outperforms Hash on HBM.
%          Exercising high bandwidth of HBM requires the participation of all cores.
\label{fig:motiv-groupby}
          }

\begin{comment} % --- ksm ---
        	Sorting achieves higher throughput, memory bandwidth,
          and scaling than hashing on HBM. Even on DRAM, sorting
          performs better than hashing. Achieving the highest
          bandwidth on HBM requires the participation of all cores.
\end{comment}

%\vspace{-2pt}		% use as needed
\label{fig:motiv}
\end{figure}

\begin{comment}
hide xeon result (we can mention)
add hash/cache (good one)

sort - square
hash - use X

line style to diff mem configs
\end{comment}

Figure~\ref{fig:motiv-groupby} compares the throughput and bandwidth of
\sort{} and \hash{} on HBM and DRAM.
%and using sort where HBM acts as a hardware-managed cache for DRAM.
The x-axis shows the number of cores. %participating in the computation.
We make the following observations.
(1)  \sort{} achieves the highest throughput and bandwidth when all cores participate.
%Although throughput of hashing scales weakly up to 32 participating cores on both HBM and DRAM, it stalls on outstanding memory requests.
(2) When parallelism is low (fewer than 16 cores),
the sequential accesses in \sort{} cannot generate enough memory traffic to
fully exercise HBM high bandwidth, exhibiting throughput similar to \sort{} on DRAM. % and achieve high throughput.
(3) HBM reverses the existing DRAM preference between
\sort{} and \hash{}. On DRAM, \sort{} is limited by
memory bandwidth and underperforms \hash{} on more than 40
cores. On HBM, \sort{} outperforms \hash{} by over 50\% for all core counts.
\hash{} experiences limited  throughput gains (10\%) from HBM, mostly
due to its sequential-access partitioning phase.
\sort{}'s advantage over \hash{} is likely to grow as HBM's bandwidth continues to scale~\cite{6757501}.
(4) HBM favors sequential-access algorithms even though they incur higher algorithmic complexity. 
%\sout{This result adds a success for sort on HBM to the long-standing ``sort or hash'' debate} ~\cite{merrettvldb1983,Albutiu:2012:MPS:2336664.2336678,sort-vs-hash-2013, sort-vs-hash-2009,polychroniou15sigmod}.

Prior work explored tradeoffs for \sort{} and \hash{} on
DRAM~\cite{sort-vs-hash-2013, sort-vs-hash-2009, 
  polychroniou15sigmod}, concluding \hash{} is best for DRAM. But
our results draw a \emph{different} conclusion for HBM -- \sort{} is best for HBM. %, even comparing to the state-of-the-art \hash{} implementation on KNL~\cite{cheng17cikm}. 
Because HBM employs a total wider bus (1024 bits vs. 384 bits for DRAM) with a wider SIMD vector (AVX-512 vs. standard AVX-256), it changes the tradeoff for software.
%\mj{figure 2 can be put side-by-side with proper graphing trick.}

% \ksm{Do we have measurements that exactly measure the amount of
%   sequential access, e.g.,  show row locality is high and the
%   prefetcher is effective?  Can we report outstanding memory requests?}

\paragraph{Why are existing engines inadequate?}
Existing engines have shortcomings that limit their efficiency on hybrid memories.  (1)
Most engines use hash tables and trees, which poorly match HBM~\cite{beam,flink,trill,streambox,sparkstreaming}.  (2) They lack
mechanisms for managing data and intermediate
results between HBM and DRAM.  Although the hardware or OS
could manage data placement~\cite{Wang:2016,sally-semantics-hybrid,Zhang:2009}, their
\emph{reactive} approaches use caches or pages, which are insufficient to manage the
complexity of stream pipelines. (3) Stream workloads may vary over
time due to periodic events, bursty events, and data resource availability.
% For example, the rate of ingress or number of records per window may vary due to periodic events or data source availability. 
Existing engines lack mechanisms for controlling 
the resultant time-varying demands for hybrid memories.
(4) With the exception of StreamBox~\cite{streambox},
most engines generate pipeline parallelism, but do not generate
sufficient total parallelism to saturate HBM
bandwidth.  % Streambox adds data parallelism within a window.

\section{Overview of \sys{}}
\label{sec:overview}

We have three system design challenges: (1) creating sequential
access in stream computations; (2) choosing which computations and data to map
to HBM's limited capacity; and (3) trading off HBM bandwidth and
limited capacity with
DRAM capacity and limited bandwidth.  To address them, we define a new smaller
extracted data structure, new primitive operations, and a new
runtime. This section overviews these components and subsequent
sections describe them in detail.
\paragraph{Dynamic record extraction}  
\sys{}
%\sout{compresses} \hym{extracts} records to essential grouping keys and record pointers
dynamically extracts needed keys and record pointers 
in a KPA data structure and operates on KPAs in HBM.
\paragraph{Sequential access streaming primitives} We implement data
  grouping primitives, which dominate stream analytics,
  with sequential-access parallel algorithms on numeric keys in KPAs. 
%These algorithms include sort, merge, join, and select.
%\sys{} only maps the grouping operations to HBM.  
The reduce primitives dereference KPA pointers sequentially, 
randomly accessing records in DRAM, and operate on bundles of records in DRAM. 
\paragraph{Plentiful parallelism} \sys{} creates computational tasks on KPA
and bundles,
producing sufficient pipeline and data parallelism to
saturate the available cores. 
\paragraph{Dynamic mapping} When \sys{} creates a grouping task, it 
  allocates or reuses a KPA in HBM. It monitors HBM capacity and
DRAM bandwidth and dynamically
balances their use by deciding where it allocates newly
created KPAs.  It never
migrates existing data.

\begin{figure}[t!]
\centering
\includegraphics[width=0.49\textwidth{}]{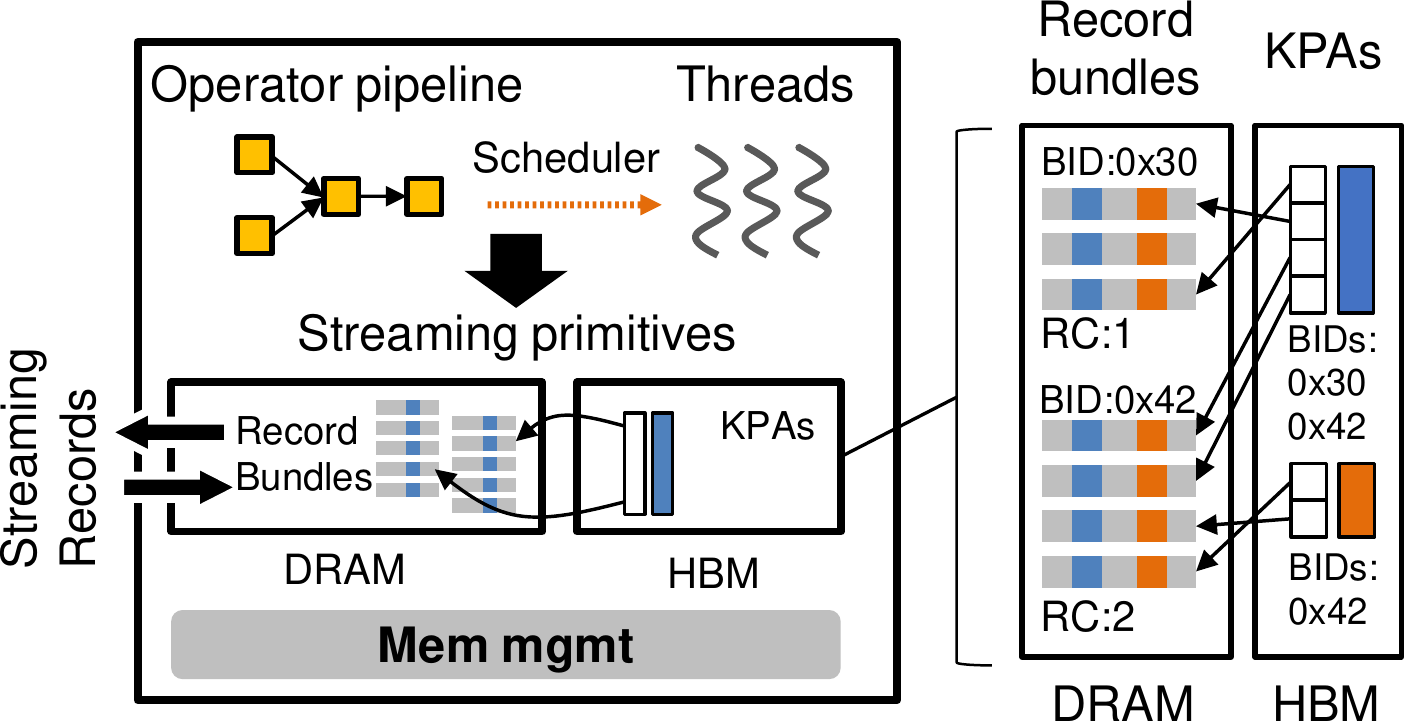} 
\caption{An overview of \sys{} using record bundles and \hbma{}s. RC: reference count; BID: bundle ID.}
\label{fig:overview}
%\vspace{-2pt}
\end{figure}

% ksm: I moved more of the detail into the short overview above, but
% not all. ;-)
\paragraph{System architecture}
\sys{} runs standalone on one machine or as multiple distributed
instances on many machines. % \st{,with each instance executing a pipeline}.  
Since our contribution is the single-machine design, we focus the remaining
discussion on one \sys{} instance.
%\mj{There should be no restriction on what to place in one machine (either slice or replica). That being said, our engine is able to be glued with any distributed stream processing way.}
Figure~\ref{fig:overview} shows how
\sys{} ingests streaming records through network sockets or RDMA and allocates them in DRAM -- in arrival order and in row format. %, i.e., one full record after the other.
%By design, \sys{} performs out-of-order stream processing as needed~\cite{OutLi2008,streambox}. It executes a pipeline correctly even when the arrival order of records differs from their event-time order.
\sys{}
dynamically manages pipeline parallelism similar to most stream
engines~\cite{flink,trill,streambox,sparkstreaming}. It further exploits data
parallelism within windows with out-of-order bundle processing,
as introduced by StreamBox~\cite{streambox}.

% !TeX root = main.tex
\section{\hbma{} and Streaming Operations}
\label{sec:kpa}

This section first
presents \hbma{} data structures (\S\ref{sec:kpa:kpa}) and primitives
(\S\ref{sec:kpa:primitives}). It then describes how \hbma{}s and the primitives
implement compound operators used by programmers (\S\ref{sec:kpa:op}), and
how \sys{} executes an entire pipeline while operating on \hbma{}s
(\S\ref{sec:kpa:pipeline}).

\subsection{KPA}
\label{sec:kpa:kpa}
To reduce capacity requirements and accelerate grouping, \sys{}
extracts \hbma{}s from DRAM and
operates on them in HBM with specialized stream
operators. Table~\ref{tab:hbma} lists the operator API.
%are \sys{}'s data structures designed for grouping data with high memory bandwidth.
KPAs are the \emph{only} data structures that \sys{} places in HBM.
A \hbma{} contains a sequence of pairs of keys and pointers pointing
to full records in DRAM, as illustrated in Figure~\ref{fig:overview}.
The keys replicate the record column required for performing the
specified grouping operation without touching the full records.
We refer to the keys in \hbma{}s as \textit{resident}. All other
columns are \textit{nonresident} keys.

One \hbma{} represents intermediate grouping results. % -- a virtual sequence of records.
%For this virtual bundle, the pointers indicate the existence of the records and the logic order among them.
The first time \sys{} encounters a grouping operation on a key $k$, it creates a
\hbma{} by extracting the specified key for each record in one bundle and creating
the pointer to the corresponding record.
To execute a subsequent grouping computation on a new key $q$, \sys{}
\emph{swaps} the \hbma{}'s resident key with the new resident key $q$ column for the
corresponding record.
After multiple rounds of grouping, one \hbma{} may contain pointers in
arbitrary order, pointing to records in arbitrary number of bundles, as
illustrated in Figure~\ref{fig:overview}. 
%\st{Each KPA data structure lists once each bundles to which it refers to efficiently reference count KPA pointers to bundles.}
Each KPA maintains a list of bundles it points to, so that the KPA can efficiently update the bundles' reference counts.
\sys{} reclaims record bundles after all the KPAs that point to them are destroyed,
as discussed in \sect{hbma:pipeline}.

%One \hbma{} resides in a contiguous region -- one memory type only (DRAM or HBM).
%They are the only data structures that \sys{} will \textit{consider} placing in HBM.

% !TeX root = main.tex

\begin{table*}[]
%\footnotesize
\vspace*{-1em}
 \begin{tabular}{@{}llll@{}}
 & \multicolumn{1}{l}{\textsf{\textbf{Primitive}}} & \multicolumn{1}{l}{\textsf{\textbf{Access}}} & \multicolumn{1}{l}{\textsf{\textbf{Description}}}  \\
 \midrule
 \parbox[t]{2mm}{\multirow{3}{*}{\rotatebox[origin=c]{90}{\small{\textsf{\textbf{Maint}}}}}} & Extract \hspace{1mm} $\mathscr{R} \rightarrow HBM(k)$ 
			& Sequential
			& Create a new \hbma{} from a record bundle. \\
 & Materialize \hspace{1mm} $\hbma(c) \rightarrow \mathscr{R}$ 
			& Random 
			& Emit a bundle of full records according to \hbma{}. \\
 & KeySwap \hspace{1mm} $\hbma(c_1) \rightarrow \hbma(c_2)$ 
			& Random
			& Replace a \hbma{}'s keys with a nonresident column. \\
 \midrule
 \parbox[t]{2mm}{\multirow{4}{*}{\rotatebox[origin=c]{90}{\small{\textsf{\textbf{Group}}}}}} 
			& Sort \hspace{1mm} $\hbma(c) \rightarrow \hbma(c)$
			& Sequential 
			& Sort the \hbma{} by resident keys \\
 & Merge $\hbma_1(c), \hbma_2(c) \rightarrow \hbma_3(c)$  
	& Sequential  
	& Merge two sorted \hbma{}s by resident keys \\
 & Join \hspace{1mm} $\hbma_1(c), \hbma_2(c) \rightarrow \mathscr{R}$  
	& Sequential 
	& Join two sorted \hbma{}s by resident keys. Emit new records. \\
 & Select \hspace{1mm} $\mathscr{R}$ or $\hbma_1(c) \rightarrow \hbma_2(c)$ 
 	& Sequential
 	& Subset a bundle as a \hbma{} with surviving key/pointer pairs.\\
 & Partition \hspace{1mm} $\hbma(c) \rightarrow \{\hbma_i(c)\}$ 
 	& Sequential
 	& Partition a \hbma{} by ranges of resident keys.\\ 	
 \midrule
 \parbox[t]{2mm}{\multirow{2}{*}{\rotatebox[origin=c]{90}{\small{\textsf{\textbf{Reduce}}}}}}  	
	& Keyed \hspace{1mm} 	$\hbma(c) \rightarrow \mathscr{R}$ 
	& Random
	& Do per-key reduction based on the resident keys. \\
 & Unkeyed	\hspace{1mm}	$\mathscr{R}_1$ or $\hbma \rightarrow \mathscr{R}_2$ 
 	& Random
	& Do reduction across all records. \\[1ex]
 \end{tabular}
\centering
%\caption{Definitions}
\caption{\hbma{} primitives. 
$\mathscr{R}$ denotes a record bundle. 
$\hbma{}(c)$ denotes a \hbma{} with resident keys from column $c$. 
%\textbf{How about ForAll primitive?}
}
\label{tab:hbma}
%\vspace*{-1em}
%\vspace{-10pt}		% use as needed
\end{table*}

\paragraph{Why one resident column?}
We enclose \emph{only one} resident column \hbma{} because this choice
greatly simplifies the implementation and reduces HBM memory consumption.
We optimize grouping algorithms for a specific data type -- key/pointer pairs,
rather than for tuples with an arbitrary column count.  Moving key/pointer
pairs and swapping keys prior to each grouping operation is much cheaper than
copying arbitrarily sized multi-column tuples.

\vspace*{-1ex}
\subsection{Streaming Operations}
\label{sec:kpa:primitives}

%\note{flatmap: is this reduction? Reduction: those producing new types (columns); those not. -- emitting \hbma{}}

\sys{} implements the streaming primitives in
Table~\ref{tab:hbma}, and the compound operators in Table~\ref{tab:operators}. The primitives fall into the following categories.
%We use $\mathscr{R}$ to denote a record bundle.
%We use $\hbma{}(c)$ to denote a \hbma{} with its resident keys from record column $c$.

%\input{tab-hbma}

\begin{myitemize}
\item
%\paragraph{Creation}
\noindent
\textbf{Maintenance primitives} convert between \hbma{}s and record
bundles and swap resident keys.
\textit{Extract} initializes the resident column by copying the key
value and initializing record pointers.
\textit{Materialize} and \textit{KeySwap} scan a \hbma{} and dereference the pointers.
% in random access.
\textit{Materialize} copies records to an output bundle in DRAM.
\textit{KeySwap} loads a nonresident column and overwrites its resident key.

\item
\noindent
\textbf{Grouping primitives} % operate on \hbma{}s with sequential memory access to exploit high memory bandwidth (\S\ref{sec:bkgnd}).
% The core grouping primitives are
\textit{Sort} and \textit{Merge} compare resident keys and rearrange key/pointer pairs within or across \hbma{}s.
Other primitives simply scan input \hbma{}s and produce output in sequential order.
% We will discuss their optimizations below.

\item
\noindent
\textbf{Reduction primitives} iterate through a bundle or \hbma{} once and produce new records.
% In doing so,
They access nonresident columns with mostly random access.
\textit{Keyed} reduction scans a \hbma{}, dereferences the \hbma{}'s pointers, locates full records, and consumes nonresident column(s).
%For instance, E.g.,
Per-key aggregation scans a sorted \hbma{} and keeps track of contiguous key ranges.
For each key range, it coalesces values from a nonresident column.
\textit{Unkeyed} reduction scans a record bundle, consumes nonresident column(s), and produces a new record bundle.
% note: it can scan hbma too

\end{myitemize}

\paragraph{Primitive Implementation}
Our design goal for primitive operations is to ensure that they all
have high parallelism and that grouping primitives produce sequential 
memory access. % We use wide vector x86 
% instructions (AVX-512) to help meet these goals~\cite{jeffers2016intel}
All primitives operate on 64-bit value key/pointer pairs.  They compare keys and based on the
comparison, move keys and the corresponding pointers.

Our
\textit{Sort} implementation is a multi-threaded merge-sort. It first splits the input \hbma{} into
$N$ chunks, sorts each chunk with a separate thread, and then merges the $N$
sorted chunks.  A thread sorts its chunk by splitting the chunk
into \emph{blocks} of 64$\times$ 64-bit integers, invoking a bitonic sort
on each block, and then performing a bitonic merge. We hand-tuned the bitonic sort and merge kernels with
AVX-512 instructions for high data parallelism. After sorting chunks,
all $N$ threads participate in pairwise merge of these chunks iteratively.  As
the count of resultant chunks drops below $N$, the threads slice chunks at key
boundaries to parallelize the task of merging fewer, but larger chunks among them.
\textit{Merge} reuses the parallel merge logic in \textit{Sort}. \textit{Join}
first sorts the input \hbma{}s  by the \emph{join} key.
It then scans them in one pass -- comparing  keys and emitting records along the way.

\paragraph{Compound Operators}
\label{sec:hbma:op}
\label{sec:kpa:op}
We  implement four common families of compound
operators  with streaming primitives and \hbma{}s.
% The four families cover most operators in popular stream analytics engines.

%\input{fig-filter}

\begin{myitemize}
\item
\noindent
\textbf{\textbf{ParDo}} is a stateless operator that applies the same function to every record,
e.g., filtering a specific column.   \sys{} implements \emph{ParDo} by scanning the input in sequential order.
If the ParDo does not produce new records (e.g., \emph{Filter} and
\emph{Sample}), \sys{} performs \textit{Selection} over \hbma{}.  When
they produce new records (e.g., \emph{FlatMap}), \sys{} performs
\textit{Reduction} and emits new records to DRAM.

\item
\noindent \textbf{Windowing} operators group records into temporal
windows using \textit{Partition} on \hbma{}.  They treat the timestamp column
as the partitioning key and window length (for fixed windows) or slide length
(for sliding windows~\cite{cql}) as the key range of each output partition.
%For fixed window, the range is the window length; for sliding window, the
%range is the slide length.
%group records by their record timestamps.

%\textbf{Selection}, e.g. filtering and sampling, scan full records (or \hbma{}, if exists) in sequence and creates a new \hbma{} containing pointers to selected records.
%No record is modified or emitted.

%\paragraph{Aggregation over all}, e.g. filtering, sampling, or flatmap, is executed by examining the full records.
%As an optimization, if \hbma{} is already created and contains resident keys accessed by the reduction, the reduction can directly execute on the \hbma{}.
%If no new record is produced, the operator produces a new \hbma{} referring to the existing record bundle;
%otherwise, the operator produces new records as a new bundle.
%Either way, the input record bundle is intact.
%arithmetic among \hbma{}s.

% !TeX root = main.tex

\begin{figure}[t!]
\centering
\includegraphics[width=0.45\textwidth{}]{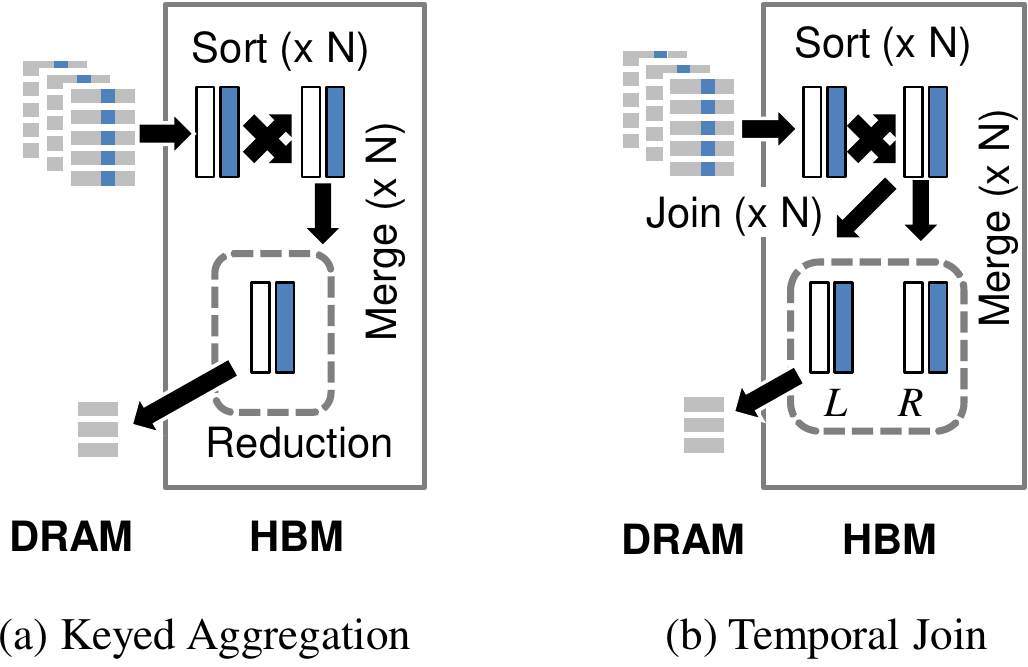} 
%\missingfigure[figwidth=6cm]{Kernel support}
%\vspace{-10pt}		% use as needed
\caption{Declarative operators implemented atop \hbma{}s}
\label{fig:op}
%\vspace{-2pt}		% use as needed
\end{figure}

\item
%\paragraph{\textbf{Keyed Aggregation}} 
\noindent
\textbf{\textbf{Keyed Aggregation}}
is a family of statefull operators that
aggregate given column(s) of the records sharing a key (e.g.,
\emph{AverageByKey} and \emph{PercentileByKey}).
\sys{} implements them using a combination of \textit{Sort} and
\textit{Reduction} primitives, as illustrated in Figure~\ref{fig:op}a.
As $N$ bundles
of records in the same window arrive, the operator extracts $N$
corresponding \hbma{}s, sorts the \hbma{}s by key, and saves the sorted
\hbma{}s as internal state for the window (shown in the
dashed-line box).  When the operator observes the window's closure by
receiving a watermark from upstream, it merges all the saved \hbma{}s by key
$k$. The result is a \hbma{}($k$) representing all records in the window
sorted by $k$. The operator then executes per-key aggregation as
out-of-\hbma{} reduction as discussed earlier.  The implementation
performs each step in parallel with all available threads.  As an
optimization, the threads perform early aggregation on individual \hbma{}s
before the window closure.

\item
\noindent 
\textbf{\textbf{Temporal Join}} takes two record streams L
and R. If two records, one in L and one in R in the same temporal
window, share a key, it emits a
new combined record.  Figure~\ref{fig:op}b shows the
implementation for R. % (for simplicity, figure illustrates input from side
                % R only).
For the $N$ input bundles in R, \sys{} extracts
their respective \hbma{}s, sorts the \hbma{}s, and performs two types of
primitives in parallel: (1) \emph{Merge}: the operator merges all the sorted
\hbma{}s by key.  The resultant \hbma{} is the window state for
R, as shown inside the dashed line box of the figure.
%The resultant two \hbma{}s are the window states for side R and L,
%respectively.  They are shown inside the dashed line box of the figure.
(2) \emph{Join} with L: in parallel with \textit{Merge}, the operator joins each of
the aforementioned sorted \hbma{} with the window state on L shown in the
dashed line box.  \sys{} concurrently performs the same procedure on
L. It uses primitive \textit{Join} on two
sorted \hbma{}($k$)s, which scans both in one pass.  The operator emits to DRAM
the resultant records, which carry the join keys and any additional columns.
\end{myitemize}

\begin{comment}
The result is a bag of \hbma{}($k$) sorted on $k$, each corresponding to a temporal window.

The operator merges each \hbma{}($k$) with the internal state on \textit{this} side, and joins it with the one on the \textit{other} side.
As described earlier, join between two sorted \hbma{}($k$)s are simply scanning both \hbma{}s once.
The operator emits the resultant records, which carries $k$ and additional columns from both sides, as full records in DRAM.
\end{comment}

%\vspace*{0em}
\subsection{Pipeline Execution Over \hbma{}s}
\label{sec:hbma:pipeline}
\label{sec:kpa:pipeline}

% \paragraph{When to build \hbma{}?}
During pipeline execution, \sys{} creates
and destroys \hbma{} and swaps resident keys dynamically. It seeks to
execute grouping operators on \hbma{} and minimize the number of accesses to nonresident columns
in DRAM.  At pipeline ingress, \sys{} ingests full records into DRAM.
Prior to executing any primitive, \sys{} examines it and transforms
the input of grouping
primitives as follows.
%\sys{} examines the operator input: if the input is \hbma{}(k), \sys{}
%executes grouping on the \hbma{}(k); if the input is \hbma{} on a different
%key, \sys{} swaps the key with $k$ before executing grouping; if the input is
%record bundle, \sys{} creates \hbma{}(k).  By doing so, \sys{} passes \hbma{}s
%among consecutive operators until an operator emits new records to DRAM.
%\note{improve}
% !TeX root = main.tex

% ref: https://tex.stackexchange.com/questions/144170/lstlistings-reference-to-line-number
\lstset{
	%language=C++,
	%basicstyle=\ttfamily\footnotesize\small,
	basicstyle=\fontsize{9}{9}\selectfont\ttfamily,
	xleftmargin=1.0ex,
	framexleftmargin=1.0ex,
	frame=tb, % top & bottom
	breaklines=true,
	captionpos=b,
%	numbers=left,
	label=list:api,
	%belowskip=0.1 \baselineskip
	}
%\begin{lstlisting}[caption={The Code, showing the use of abstractions. The hints will be elaborated in Section~\ref{sec:mm:placement}},escapechar=^]
\begin{lstlisting}[escapechar=^]
/* X: input (a KPA or a bundle) */
/* c: column containing grouping key */
X = IsKPA(X) ? X : Extract(X)
if ResidentColumn of X != c 
  KeySwap(X, c)
Execute grouping on X  
\end{lstlisting}

\sys{} applies a set of optimizations to further reduce the number of DRAM accesses.
(1)
%Inter-primitive coalescing:
It coalesces adjacent \textit{Materialize} and \textit{Extract} primitives
to exploit data locality. As a primitive emits new records to DRAM, it
simultaneously extracts the \hbma{} records required by the next
operator in the pipeline. % This avoiding accessing the full records again in
%In-place update of \hbma{}. A \hbma{} is immutable for most of the time.
(2) It
updates \hbma{}'s resident keys in place, and writes back dirty
keys to the corresponding nonresident column as needed for future \textit{KeySwap}
and \textit{Materialize} operations.  (3) It avoids extracting %short
 records that
contain fewer than three columns, which are already compact.

% !TeX root = main.tex

\begin{wrapfigure}{R}{0.24\textwidth}

%\begin{figure}
\centering
\vspace{-15pt}		% use as needed
\includegraphics[width=0.24\textwidth]{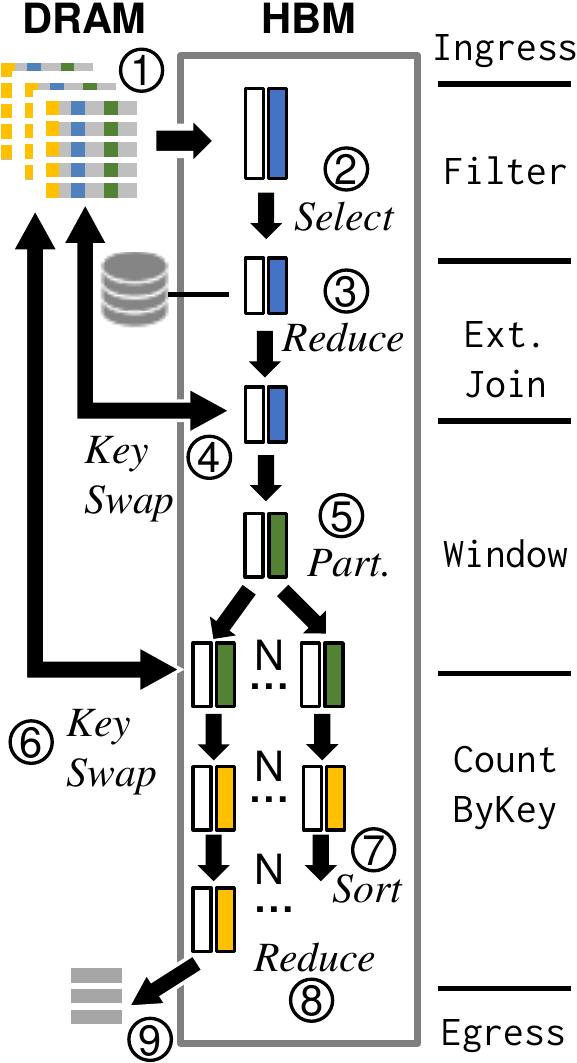}
%\vspace{-10pt}		% use as needed
%\caption{Inexpensive IoT devices could help democratize \ib{}. The IoT device: A \$10 Raspberry Pi Zero converted to a wall plug for sensing} 
%\caption{A \$10 IoT device (Raspberry Pi Zero W), packed as a wall plug, are ideal for sensing and controlling tasks in \ib{}s. Photo: raspberrypi.org} 
\caption{Pipeline execution on \hbma{}s for YSB~\cite{ycsb}. Declarative operators shown on right.} 
%Example in (b)~\cite{smart-power-plugs}}
\label{fig:op-ycsb}
\vspace{-5pt}		% use as needed
%\end{figure}

\end{wrapfigure}

% \paragraph{A mini example -- YCSB}
\paragraph{Example} We use YSB~\cite{ycsb} in
Figure~\ref{fig:pipeline-stacked:pipeline} to show  pipeline execution. We omit
\emph{Projection}, since \sys{} stores results in DRAM. % implements as \code{nop} to minimize record access.
Figure~\ref{fig:op-ycsb} shows the engine ingesting record bundles to DRAM
\circlew{1}.  Filter, the first operator, scans and selects records based on
column $ad\_type$, producing \hbma{}($ad\_id$) \circlew{2}.
% where $ad\_id$ is the key for the next operator.
%Join consumes the incoming \hbma{}(k) and emits full records to DRAM (since it
%produces new records); At the same time, \sys{} creates \hbma{}() on event
%time of these records.
\emph{External Join} (different from temporal join) 
scans the KPA and updates the
resident keys $ad\_id$ in place with $camp\_id$ loaded from an external
key-value store \circlew{3}, which is a small table in HBM. The operator writes back $camp\_id$ to full
records and swaps in timestamps $t$ \circlew{4}, resulting in \hbma{}($t$).
Operator \emph{Window} partitions the \hbma{} by $t$ \circlew{5}.  \emph{Keyed Aggregation}
swaps in the grouping key $camp\_id$ \circlew{6}, sorts the resultant
\hbma{}($camp\_id$) \circlew{7}, and runs reduction on \hbma{}($camp\_id$) to
count per-key records \circlew{8}.  It emits per-window, per-key record counts
% i.e. counts of per-campaign ads,
as new records to DRAM \circlew{9}.

% !TeX root = main.tex

%\section{Parallelism Cap}
%\section{Coupling Placement And Scheduling}
%\section{XXX -- Capacity and Bandwidth Constraints}
%\section{Balancing Memory Pressure}
\section{Dynamically Managing Hybrid Memory}

% As described in \sect{bkgnd}, the tension between large stream data and HBM's
% limited capacity is fundamental.
In spite of the compactness of \hbma{}s representation,
HBM still cannot hold all the \hbma{}s at once.
%In case of HBM is able to hold only a fraction of all the \hbma{}s, HBM may
%not be able to hold all the \hbma{}s created in processing large streams.
\sys{} manages \textit{which} new \hbma{}s to place on \textit{what} type
of memory by addressing the following two concerns.

\begin{myenumerate}
\item \textit{Balancing demand}.  \sys{} balances the aggregated
  demand for limited HBM capacity and DRAM bandwidth to
  prevent either from becoming a bottleneck.

\item \textit{Managing performance}.  As \sys{} dynamically
schedules a computation, it optimizes for the access pattern,
parallelism, and contribution to the critical path by where it
allocates the KPA for the computation.
%Depending on pipeline schedule, stream computations show different performance
%expectation and parallelism in accessing the \hbma{}s.
\sys{} prioritizes creating \hbma{} in HBM for aggregation operations on the
critical path to pipeline output. When work is on the critical path,
it further prioritizes increasing parallelism
and throughput for these operations versus \hbma{} that are processing
early arriving records. We mark bundles an \emph{urgent}  on the critical path with a \emph{performance impact tag}, as described below.
%In the YCSB example, the \hbma{}s used for window aggregation have higher
%performance impact.  It is therefore vital for the placement of new \hbma{}s
%to consider different performance expectations, which is scheduling knowledge
%possessed by \sys{} runtime.
\end{myenumerate}

\sys{} monitors HBM capacity and DRAM bandwidth and trades them off dynamically.
For individual \hbma{} allocations, \sys{} further considers the
critical path. % computational performance demands.
\sys{} does not migrate existing \hbma{}s, which are ephemeral, unlike other software systems
for hybrid memory~\cite{Wang:2016,sally-semantics-hybrid,Zhang:2009}.

%Unlike many prior systems \note{cite}, \sys{} refrains from migrating existing memory objects.
%This is because streaming objects inside a pipeline are frequently allocated/freed and are ephemeral.
%Migrating them will lead to substantial performance loss, as we will show in \sect{eval}.

% !TeX root = main.tex

\begin{figure}[t!]
\centering
\includegraphics[width=0.48\textwidth{}]{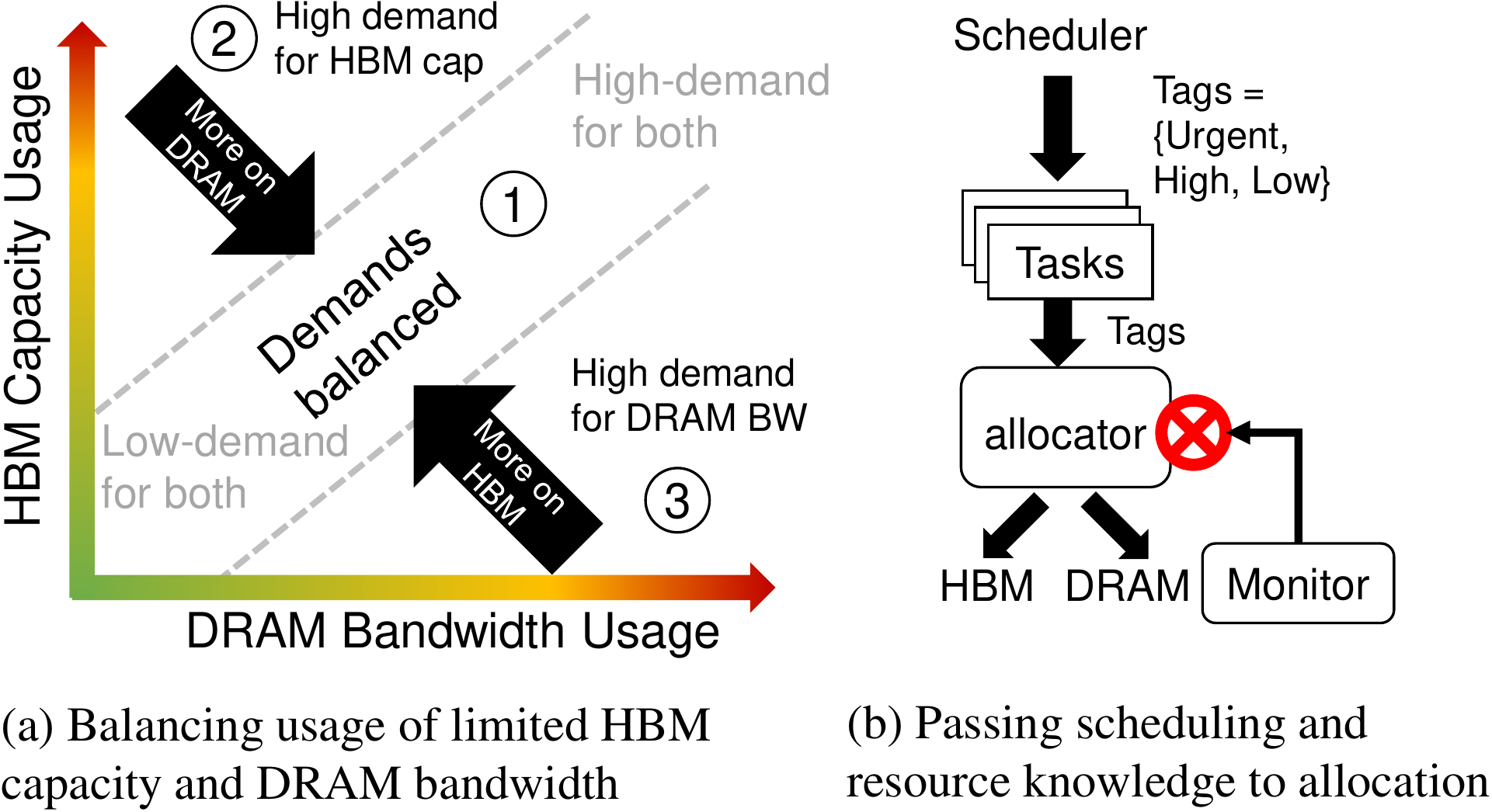} 
%\missingfigure[figwidth=6cm]{Kernel support}
%\vspace{-10pt}		% use as needed
%\caption{\sys{} dynamically decides new \hbma{} placement so it balances the usage of two limited resources in hybrid memory and differentiates \hbma{} performance demands.}
\caption{\sys{} dynamically manages hybrid memory}
\label{fig:pressure}
%\vspace{-2pt}		% use as needed
\end{figure}

\paragraph{Dynamically Balancing Memory Demand}
%We next describe our principle for balancing the memory demands.
%Figure~\ref{fig:pressure} shows how \sys{} controls its state in the space of two limiting factors -- HBM capacity and DRAM bandwidth.
Figure~\ref{fig:pressure} plots \sys{}'s state space.  \sys{} strives to
operate in the diagonal zone \circlew{1}, where limiting capacity and bandwidth demands are balanced.
%In this zone, when both demands are low (e.g. low ingestion rate, fewer
%records per window), the system operates in the bottom-left area of the state
%space; as both demands increase from low to high, the state moves from the
%bottom-left zone towards the top-right zone.
If both capacity and bandwidth reach their limit, \sys{} operates in the top-right
corner in zone \circlew{1}, while throttling the number of concurrent threads
working on DRAM to avoid over-subscribing  bandwidth and wasting cores, and preventing back pressure on ingestion.

%As the HBM usage increases (e.g. operator execution cannot free \hbma{}s in
%time) or the DRAM bandwidth usage increases (e.g. \note{fix}), \sys{} is
%moving away from the diagonal zone towards zone \circled{2} or \circled{3},
%respectively.
When the system becomes imbalanced, the state moves away from
zone \circlew{1} to \circlew{2} or \circlew{3}.  Example causes include
additional tasks spawned for DRAM bundles which stress DRAM
bandwidth,
%that earlier operator tasks take long to process \hbma{}s on DRAM after new
%tasks start, increasing DRAM bandwidth usage;
and delayed watermarks that postpone window closure which stresses HBM capacity.
%and that a sudden increase in ingestion rate bumps up HBM capacity usage (e.g.
%\note{fix}).
If left uncontrolled, such imbalance will lead to performance degradations. When
HBM is full, all future \hbma{}s regardless of their performance
impact tag are
forced to spill to DRAM.
%If HBM is completely filled, all future \hbma{}s, including the ones expecting
%high parallelism/performance, are forced to spill to DRAM;
When DRAM bandwidth is fully saturated, additional parallelism on DRAM wastes
cores. % adds no performance.  Such degradation could have been avoided with
%available memory resource.

At runtime, \sys{} balances resources by tuning a global \textit{demand balance
knob} as shown in Figure~\ref{fig:pressure}. \sys{} gradually changes
the fraction of the \textit{new} \hbma{} allocations on HBM or DRAM, and pushes
its state back to the diagonal zone.  % We will present the demand balance knob details in
% \sect{mm:impl}.
%In rare cases, the system is limited because there is no more HBM capacity and no more DRAM bandwidth. We control ingestion by using a pull model for RDMA …
In rare cases, there is no more HBM capacity and no more DRAM bandwidth because the data ingestion rate is too high. To address this issue, \sys{} dynamically starts or stops pulling data from data source according to current resource utilization.

%\subsection{Instantiation}
\label{sec:mm:impl}

%\note{need extra details: can DRAM run out? what if both resource full? stall new tasks? etc.}

%\sys{} instantiates the above designs as follows.

\paragraph{Performance impact tags} 
To identify the critical path, \sys{} maintains a global target watermark, which indicates the next window to close. 
\sys{} deems any records with timestamps earlier than the target watermark on the critical path.
When creating a task, the \sys{}
scheduler tags it with one of three coarse-grained \emph{impact} tags
based on when the window that contains the data for this task will be
externalized. Windows are externalized based on their
record-time order. %, which we find sufficient in
%practice.
(1) \textit{Urgent} is for tasks on the critical path of pipeline
output.  Examples include the last task in a pipeline that aggregates the current window's internal
state.  (2) \textit{High} is for tasks
on younger windows (i.e., windows with earlier record time), for which results
will be externalized in the \textit{near} future, say one or two
windows in the future. (3) \textit{Low} is for tasks
on even younger windows, for which results will be externalized in the
\textit{far} future.

\begin{comment} \begin{itemize} \item \textit{Urgent} is for tasks on the
critical path of pipeline output.  Examples include a task for aggregating the
current window's internal state upon the closure of this window.  \item
\textit{High} is for tasks working on ``older'' windows (i.e. windows with
earlier record time), for which results will be externalized in the
\textit{near} future.  This is because a pipeline externalizes its windows in
record-time order.  \item \textit{Low} is for tasks working on younger windows,
for which results will be externalized in the \textit{far} future.
\end{itemize}
\end{comment}

\paragraph{Demand balance knob} We implement a demand balance knob as a
global vector of two scalar values $\{k_{low}, k_{high}\}$, each in the range
of $[0,1]$.  $k_{low}$ and $k_{high}$ define the probabilities for \sys{} to allocate
\hbma{}s on HBM for Low and High tasks correspondingly.  \textit{Urgent} tasks
always allocate \hbma{}s from a small reserved pool of HBM.  The knob in
conjunction with each \hbma{} allocation's performance impact tag determines the
\hbma{} placement as follows.
%\sys{} evaluate such decisions for every allocation based on the
%\textit{current} knob values.

% !TeX root = main.tex

% ref: https://tex.stackexchange.com/questions/144170/lstlistings-reference-to-line-number
\lstset{
	%language=C++,
	%basicstyle=\ttfamily\footnotesize\small,
	basicstyle=\fontsize{9}{9}\selectfont\ttfamily,
	xleftmargin=1.0ex,
	framexleftmargin=1.0ex,
	frame=tb, % top & bottom
	breaklines=true,
	captionpos=b,
%	numbers=left,
	label=list:api,
	%belowskip=-0.0 \baselineskip
	}
%\begin{lstlisting}[caption={The Code, showing the use of abstractions. The hints will be elaborated in Section~\ref{sec:mm:placement}},escapechar=^]
\begin{lstlisting}[escapechar=^]
/* to choose memory type to be M */
switch (alloc_perf_tag)
case Urgent:
  M = HBM
case High:
  M = random(0,1) < k_high ? HBM : DRAM
case Low:
  M = random(0,1) < k_low ? HBM : DRAM
allocate on M
\end{lstlisting}

\sys{} refreshes the knob values every time it samples the monitored resources.
It changes the knob values in small increments $\Delta$ for controlling
future HBM allocations.
% based on the algorithm listed in XXX \note{need a
% piece of code In doing so, it avoids sudden impacts on near-term performance:
%In doing so, it first attempts to change $k_{low}$ as the value long-term
%performance impact;
To balance memory demand it first considers changing $k_{low}$; if $k_{low}$ already reaches an extreme
(0 or 1), \sys{} considers changing $k_{high}$ if the pipeline's current output delay
still has enough headroom (10\%) below the target delay.
We set the initial values of $k_{high}$ and $k_{low}$ to 1, and set $\Delta$ to 0.05.

\subsection{Memory management and resource monitoring}
% \paragraph{Memory allocation and resource monitoring} 
\sys{} manages
HBM memory with a custom slab allocator on top of a memory pool with
%fixed-sized elements of size XX, YY, ... \note{FIXME}, tuned to typical KPA sizes.
different fixed-sized elements, tuned to typical KPA sizes, full record bundle sizes, and window sizes.
The allocator tracks the amount of free
memory. \sys{} measures DRAM bandwidth usage with Intel's processor
counter monitor library~\cite{intel-pcm}. \sys{} samples both metrics
at 10$\;$ms intervals, which are sufficient for our analytic
pipelines that target sub-second output delays.

%\paragraph{Managing record bundles}
%\paragraph{Reclaiming bundles} 
By design,
%\sys{} executes all grouping on \hbma{}s and
\sys{} never modifies a bundle by adding, deleting, or reordering
records.  After multiple rounds of grouping, all records in a bundle may be
dead (unreferenced) or alive but referenced by different \hbma{}s.
\sys{} reclaims a bundle when no \hbma{} refers to any record in
the bundle using reference counts (RC). On the \hbma{} side, each \hbma{}
maintains one reference for each source bundle to which any record in
the \hbma{} points. On the bundle side, each bundle stores a
reference count (RC) tracking how many \hbma{}s link to it. When \sys{}
extracts a new \hbma{} ($\mathscr{R}\rightarrow \hbma{}$), it adds a
link pointing to $\mathscr{R}$ %; $\mathscr{R}$ 
if one does not exist and increments the reference count.
When it destroys a \hbma{}, it follows all the \hbma{}'s links to locate
source bundles and decrements their reference counts.  When merging or
partitioning \hbma{}s, the output \hbma{}(s) inherits the input \hbma{}s' links
to source bundles, and increments reference counts at all source
bundles.  When the reference count of a record bundle drops to zero, \sys{}
destroys the bundle.

% !TeX root = main.tex

\section{Implementation and Methodology}
\label{sec:impl}

We implement \sys{} in C++ atop StreamBox, an open-source research analytics engine~\cite{streambox-code,streambox}. 
%\sys{} has 61K lines of code in total, which includes 23K lines of StreamBox and 38K lines of new implementations of this work. 
\sys{} has 61K lines of code, of which 38K lines are new for this work.
\sys{} reuses StreamBox's work tracking and task scheduling, which
generate task and pipeline parallelism. We introduce new operator implementations and
novel management of hybrid memory, replacing all of the StreamBox
operators and enhancing the runtime, as described in the previous
sections. The current implementation supports numerical data, which is very common in data analytics~\cite{tersecades}. %We will support more data types in future work.
%A key parameter in \sys{} is bundle size, which we empirically determine as 100$\;$K records. % hym: not necessary

%Key parameters in \sys{} are window and bundle size. We use a default
%window size of 10$\;$M records that spans one second of event time and bundle size of 100$\;$K records.

% While StreamBox heavily uses hash tables and trees which mismatch HBM, 
% \sys{} introduces completely different operator algorithms and memory management, which is crucial to exploiting the hybrid memory. 

\paragraph{Benchmarks}
We use 10 benchmarks with a default window size of 10$\;$M records that spans one second of event time. 
One is YSB, a widely used streaming
benchmark~\cite{ycsb-flink,ycsb-spark,drizzle}.
YSB processes input records with seven columns, for which we use numerical values rather than JSON strings. 
%hym: we will discuss data types and parsing in S7.4, so delete these sentences here.
%due to the lack of publicly available JSON parsers with state-of-the-art performance~\cite{mison}. 
%Note that production systems are carefully engineered such that parsing is not a bottleneck, e.g., Google Dataflow is using Protocol Buffers~\cite{protoco-buffers}, which has very minor overhead.
Figure~\ref{fig:pipeline-stacked:pipeline} shows its pipeline.  
%\note{generate data} 

%\note{we are robust to key skewness. mention record size? columns...}
%(1) {\bf TopVperK} finds the top 10 percent of values for each key in every window. (2) {\bf SumVperK} calculates the sum of values for each key in every window. (3) {\bf MedVperK} finds the medium of values for each key in every window. (4) {\bf AvgVperK} calculates the average of values for each key in every window. (5) {\bf UniqVperK} counts the number of unique values for each key in every window. (6) {\bf JoinByK} takes two input streams and join by keys in every window. (7) {\bf JoinFilterByK} takes two input streams, calculate the average of values for keys on one stream in every window, and use the average value to filter on another streams. (8) 

We also use nine benchmarks with a mixture of widely tested, simple pipelines (1--8) and one complex pipeline (9).
All benchmarks process input records with three columns -- keys, values, and timestamps, except that input records for benchmark 8 and 9 contain one extra column for secondary keys. 
%All columns contain 64-bit integers. %hym: mentioned below
(1) {\bf TopK Per Key}
groups records based on a key column and identifies the top K largest values for each key in each window. %We used randomly generated keys and values.
(2) {\bf Windowed Sum Per Key} 
aggregates input values for every key per window. %within each window.
(3) {\bf Windowed Median Per Key} 
%\ksm{I am pretty sure this is the mathmatical median function (medium is not a function)? Fix in the graphs please.}
calculates the median value for each key per window. %in every window. 
(4) {\bf Windowed Average Per Key}
calculates the average of all values for each key per window. %in every window. 
(5) {\bf Windowed Average All}
calculates the average of all values per window. %every window. 
(6) {\bf Unique Count Per Key} 
counts unique values for each key per window. %in every window.
(7) {\bf Temporal Join} joins two input streams by keys per window. %in every window. 
(8) {\bf Windowed Filter} takes two input streams, calculates the value average on one stream per window, and uses the average to filter the key of the other stream.
(9) {\bf Power Grid} is 
derived from a public challenge~\cite{DEBS14_power}. It finds houses with
the most high-power plugs. Ingesting a stream of per-plug
power samples, it calculates the average power of each
plug in a window and the average power over all plugs in
all houses in the window. Then, for each house, it counts
the number of plugs that have higher load than average.
Finally, it emits the houses that have most high-power
plugs in the window. 
%\ksm{How do we generate correct streaming data for it?}

For YSB, we generate random input following the benchmark directions~\cite{ycsb}.
For Power Grid, we replay the input data from the benchmark~\cite{DEBS14_power}. 
For other benchmarks, we generate input records with columns as 64-bit random integers. 
Note that our grouping primitives, e.g. sort and merge, are insensitive to key skewness~\cite{Albutiu:2012:MPS:2336664.2336678}. 
%The input records for YCSB have seven columns; 
%those for Windowed Join Filter and Power Grid have four columns;
%and those for all others have three columns. 
%For all other benchmarks, we generate records with 3 keys (the smallest size that we compress into KPAs). 
%\ksm{This has to be true, right?} \note{Right. Note YCSB has 7 keys}

\paragraph{Hardware platform}
We implement \sys{} on KNL~\cite{jeffers2016intel}, a manycore
machine with hybrid HBM/DRAM memory. 
Compared to the standard DDR4 DRAM on the machine, the 3D-stacked HBM
DRAM offers 5$\times$ higher bandwidth with 20\% longer latency.
The machine has 64 cores with 4-way simultaneous multithreading for a total of 256 hyper-threads. 
We launch one thread per core as we find out this configuration outperforms two or four hyper-threads per core
due to the number of outstanding memory requests supported by each core.
The ISA includes AVX-512, Intel's wide vector instructions.
We set BIOS to configure HBM and DRAM in \emph{flat} mode,
where both memories appear fully addressable to \sys{}.
We also compare to \emph{cache} mode, where HBM is a hardware-managed
last-level cache in front of the DDR4 DRAM.
Table~\ref{tab:plat} summarizes the KNL hardware and
a 56-core Intel Xeon server (X56) used in evaluation for comparisons. 

% !TeX root = main.tex

\begin{table}[t!]
\centering
%\vspace{1pt}
\includegraphics[width=0.48\textwidth{}]{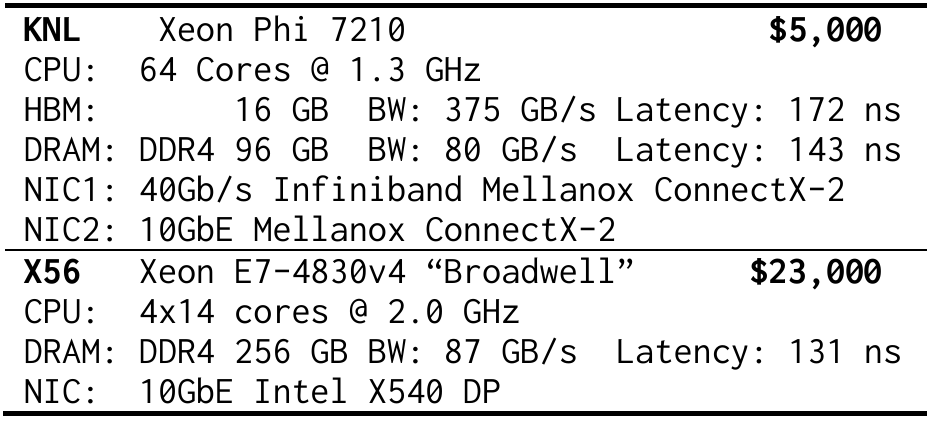}
\vspace*{-15pt}
\caption{KNL and Xeon Hardware used in evaluation}
\label{tab:plat}
%\vspace{-10pt}		% use as needed
\end{table}

\paragraph{Data ingress}
We use a separate machine (an i7-4790 with 16 GB DDR4 DRAM) called Sender to
generate input streams.  To create sufficient ingestion bandwidth, we
connect Sender to KNL using RDMA over 40$\;$Gb/s Infiniband.  
With RDMA ingestion, \sys{} on KNL pre-allocates a pool of input record bundles.
To ingest bundles, \sys{} informs Sender of the bundle addresses and 
then polls for a notification which signals bundle delivery 
from Sender.  
%During this interaction, both sides batch bundle addresses to amortize the synchronization delay.
%
%To explore sensitivity to ingestion bandwidth and
To compare \sys{} with commodity engines that do not support RDMA
ingestion, we also deliver input over our available
10$\;$Gb/s Ethernet using the popular, fast ZeroMQ transport~\cite{zmq}.  
With ZeroMQ ingestion, the engine copies
incoming records from network messages and creates record bundles
in DRAM.  

\section{Evaluation}
\label{sec:eval}

We first show \sys{} outperforms Apache Flink~\cite{flink} on YSB.
%, a state-of-the-art streaming engine, on the YCSB benchmark.  
We then evaluate \sys{} on the other  benchmarks, where it achieves high throughput by exploiting high memory bandwidth. 
We demonstrate that the key design features, \hbma{} and dynamically balancing memory and performance demands, are essentially to achieving high throughput.

%We then show \sys{} outperforms using HBM as a hardware-managed cache and the capability of balancing memory demands. 

\subsection{Comparing to Existing Engines}

\paragraph{Comparing to Flink on YSB}
We compare to Apache Flink (1.4.0)~\cite{flink}, a popular stream analytics engine known for its good single-node performance on the YSB benchmark described in Section~\ref{sec:impl}. 
To compare fairly, 
we configure the systems as follows.
(1) Both \sys{} and Flink ingest data using ZeroMQ transport over 10$\;$Gb/s Ethernet, since Flink's default, Kafka, is not fast enough and it does not ingest data over RDMA.
(2) The Sender generates records of numerical values rather than JSON strings. % due to the lack of publicly available JSON parsers with state-of-the-art performance~\cite{mison}.
We run Flink on KNL by configuring HBM and DRAM in cache mode, so that Flink transparently uses the hybrid memory.  We also compare on the high end Xeon server (X56) from Table~\ref{tab:plat} because Flink targets such systems. 
%as commodity JSON parsers are known to bottleneck stream analytics despite of much faster algorithms~\cite{mison}. 
We set the same target egress delay (1 second) for both engines.

\begin{figure}[t!]
\centering
   \begin{subfigure}[b]{0.45\textwidth}
   \includegraphics[width=1\linewidth]{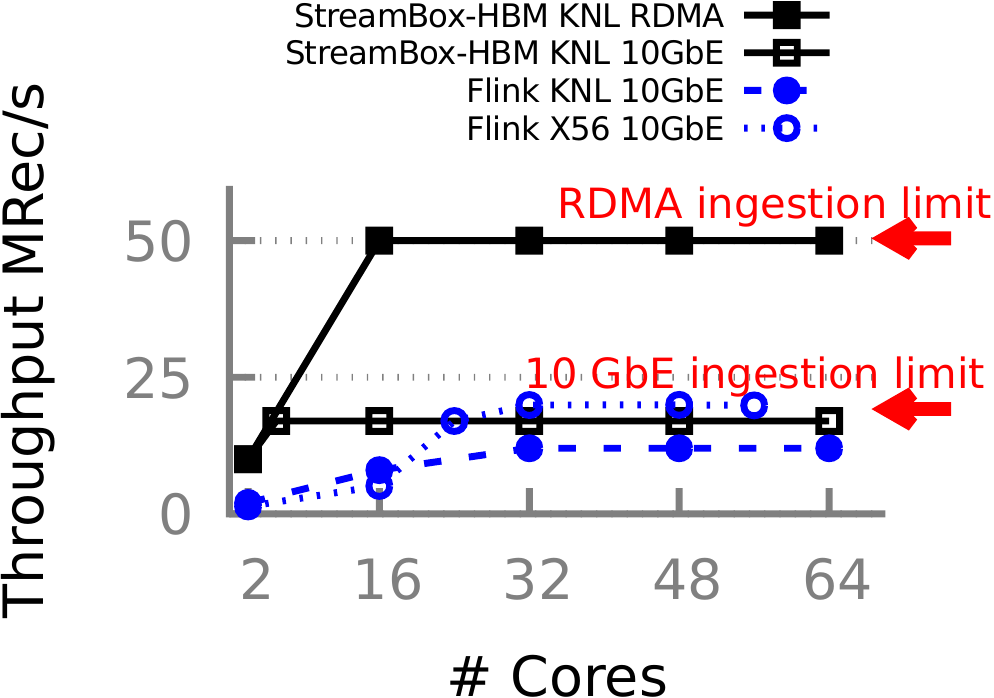}
   \caption{Input throughput under 1-second target delay. Note: X56's 10GbE NIC is slightly faster than that on KNL.}
   % Benchmark: cpp drawing tutorial 1
	\vspace{2mm}
   \label{fig:motiv:spatial} 
	\end{subfigure}
	\begin{subfigure}[b]{0.45\textwidth}
	   \includegraphics[width=1\linewidth]{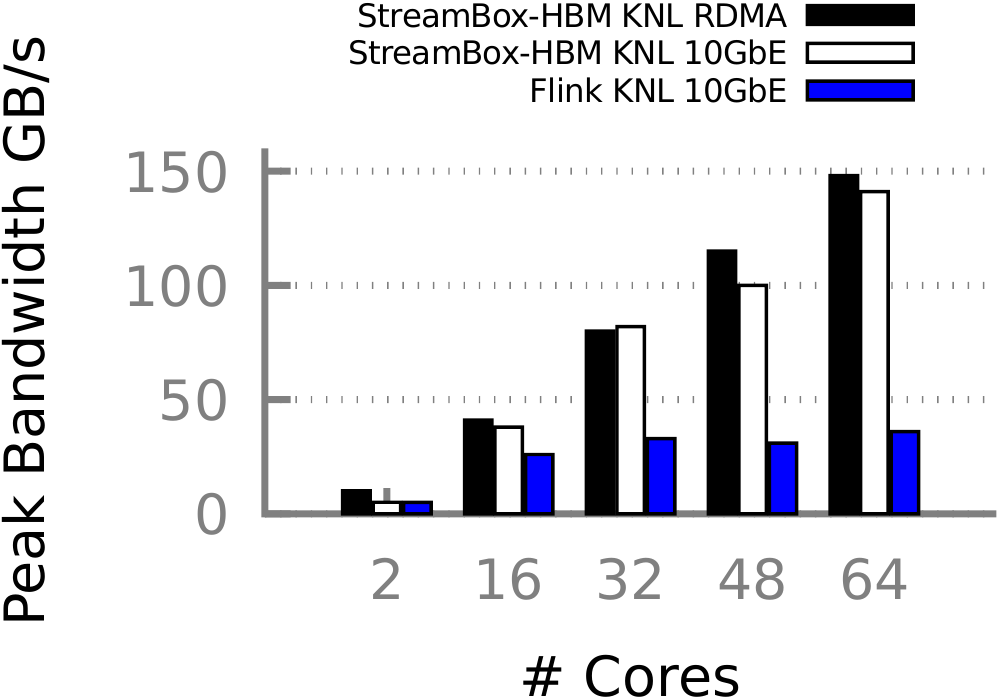}
	   %\caption{Peak memory bandwidth utilization}
	   \caption{Peak memory bandwidth usage of HBM}
		\vspace{2mm}
	\end{subfigure}
\caption{\sys{} achieves much higher throughput and memory bandwidth usage than Flink, quickly saturating IO hardware.
Legend format: ``Engine Machine IO''.
Benchmark: YSB~\cite{ycsb}}
%\vspace{-2pt}		% use as needed
%\label{fig:cmp_flink}
\label{fig:exp-ycsb}
\end{figure}

%\paragraph{Results}
Figure~\ref{fig:exp-ycsb} shows throughput (a) and peak bandwidth (b) of YSB as a function of hardware parallelism (cores). \sys{}  achieves much higher throughput than Flink on KNL.  
 It also achieves much higher per-dollar throughput on KNL than Flink running on X56, 
because KNL cost is \$5,000, 4.6$\times$ lower than X56 at \$23,000.
Figure~\ref{fig:exp-ycsb} shows when both engines ingest data over 10$\;$Gb/s Ethernet on KNL, \sys{} maximizes the I/O throughput with 5 cores while Flink cannot saturate the I/O even with all 64 cores. 
% 5 core: 17GB/s  vs   64 cores: 12GB/s. we choose these points because these are the points before they hit IO bottleneck.
By comparing these two operating points, \sys{} shows 18$\times$ per core throughput than Flink.
% inside the pipeline with more threads. 
%as usage of \sys{} goes higher as more cores are in use to do grouping and closing windows in parallel, but the throughput does not go up because \sys{} is limited by I/O, which means it not useful to have more cores with limited I/O. 
% We further run Flink on a high-end Xeon server ``56CM'' (see Table~\ref{tab:plat} for its hardware details). 
On X56, Flink saturates the 10 Gb/s Ethernet I/O when using 32 of 56 cores. 
%By comparing such throughput of \sys{} to running Flink on a high-end Xeon server (see Table~\ref{tab:plat} for details), we show that the throughput of \sys{} the same throughput on XXX cores of the \$5,000 \knl{} as Flink achieves 
% \ksm{?? The figure shows that HyStream does not saturate the bandwidth at 5 cores; it is still increasing bandwidth usage at 64 cores, but it has no impact on throughput. Why is that?} 
%\note{This figure shows the PEAK bandwidth utilizations, which come from parallel merge in closing a window. The peak bandwidth is decided by the parallelism, so it will go up as the more cores are in use, no matter how fast the data comes.}
As shown in Figure~\ref{fig:exp-ycsb}b, when \sys{} saturates its ingestion I/O, adding cores will further increase the peak memory bandwidth usage which results from \sys{} executing grouping computations with higher parallelism. 
%\sout{However, t} 
This parallelism does not increase the overall pipeline throughput which is bottlenecked by ingestion, but it reduces the pipeline's latency by closing a window faster. 
Once we replace \sys{}'s 10$\;$ Gb/s Ethernet ingestion with 40 Gb/s RDMA, its throughput further improves by 2.9$\times$  (saturating the I/O with 16 cores), leading to 4.1$\times$ higher machine throughput than Flink. Overall, \sys{} achieves 18$\times$ higher per core throughput than Flink.% before saturating I/O.
% !TeX root = main.tex

%\begin{figure}
%    \centering
%    \begin{subfigure}[b]{0.2\textwidth}
%        \includegraphics[width=\textwidth]{figs/plot/cmp_flink/cmp_bw_tput}
%        \caption{\wc{}}
%        \label{fig:gull}
%    \end{subfigure}
%    ~ %add desired spacing between images, e. g. ~, \quad, \qquad, \hfill etc. 
%      %(or a blank line to force the subfigure onto a new line)
%    \begin{subfigure}[b]{0.2\textwidth}
%        \includegraphics[width=\textwidth]{figs/plot/cmp_flink/cmp_flink_old/cmp_bw}
%        \caption{\netmon{}}
%        \label{fig:tiger}
%    \end{subfigure}
%    \caption{\sys{} achieves high throughput and uses HBM bandwidth more efficiently. }\label{fig:cmp_flink}
%\end{figure}

%\begin{figure}
%    \centering
%	\includegraphics[width=0.45\textwidth{}]{figs/plot/cmp_flink/cmp_bw_tput.pdf}
%    \caption{\sys{} uses HBM bandwidth more efficiently. \textbf{XXX: ZMQ->10GBE}}\label{fig:cmp_flink_bw}
%\end{figure}
%
%\begin{figure}
%    \centering 
%	\includegraphics[width=0.45\textwidth{}]{figs/plot/cmp_flink/cmp_flink_old/cmp_bw}
%    \caption{\sys{} achieves high throughput. \textbf{ZMQ->10GBE}}\label{fig:cmp_flink_tput}
%\end{figure}

\paragraph{Qualitative comparisons}
Other engines, e.g., Spark, and Storm, report lower or comparable performance to Flink, with at most tens of millions of records/sec per machine~\cite{esper,streambox,oraclecep,streambase,toshniwal2014storm,sparkstreaming}. 
None reports 110$\;$M records/sec on one machine as \sys{} does (shown below). 
%We expect \sys{} outperforms them by a large margin. 
%A few engines are specifically optimized for single-node performance~\cite{oraclecep,streambase,esper}: 
%with up to 16 cores, they achieve throughput up to a few million records/sec. 
%None reports to scale beyond 32 cores and achieves 110M records/sec on one machine as \sys{} does. 
Executing on a 16-core CPU and a high-end (Quadro K500) GPU, SABER~\cite{saber} reports 30$\;$M records/sec on a benchmark similar to Windowed Average, 
%which Section~\ref{sec:eval:tput} shows is 4$\times$ lower than \sys{}.
which is 4$\times$ lower than \sys{} as shown in Section~\ref{sec:eval:tput}.
On a 24-core Xeon server, which has much higher core frequency than KNL, Tersecades~\cite{tersecades}, a highly optimized version of Trill~\cite{trill}, achieves 49$\;$M records/sec on the same Windowed Average benchmark; compared to it, \sys{} achieves 2.3$\times$ higher machine throughput and 3.5$\times$ higher per core throughput before saturating the I/O.
%Since we build \sys{} atop StreamBox~\cite{streambox}, we will experimentally compare \sys{} to StreamBox in Section~\ref{sec:eval:valid} below.
In summary, \sys{} achieves much higher single-node performance than existing streaming engines. 
% In the future, other engines could include our mechanisms for harnessing HBM and DRAM.

% !TeX root = main.tex

\begin{figure*}
    \centering
    \begin{subfigure}[b]{0.3\textwidth}
        \includegraphics[width=\textwidth]{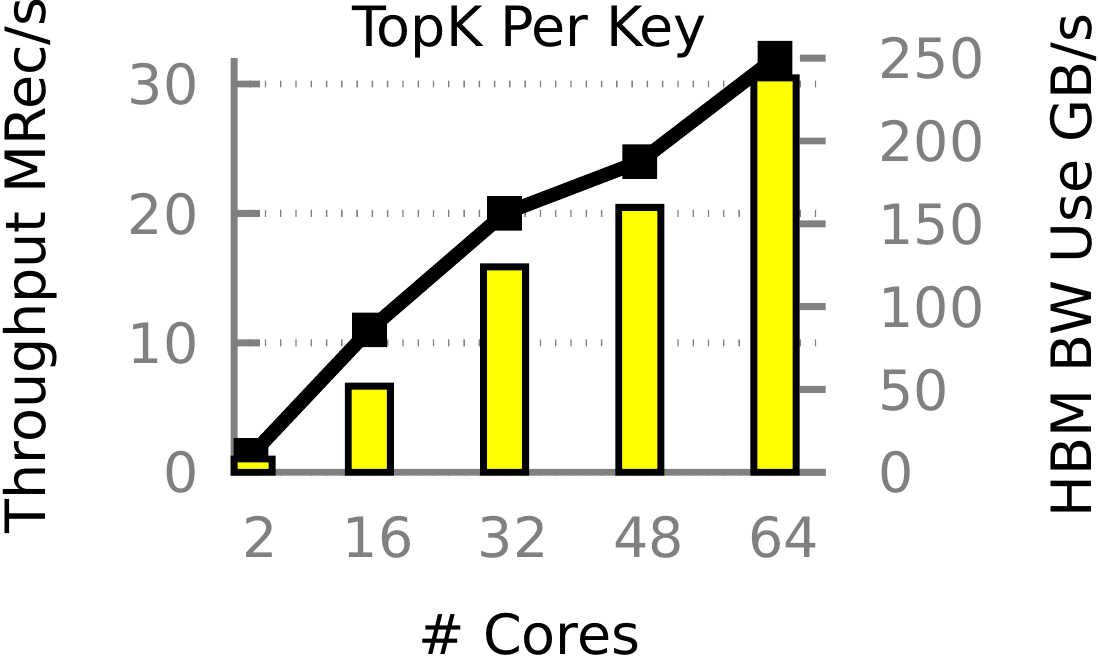}
%        \caption{wingrep \makalu{}}
        \label{fig:gull}
    \end{subfigure}
    ~ %add desired spacing between images, e. g. ~, \quad, \qquad, \hfill etc. 
    \begin{subfigure}[b]{0.3\textwidth}
\includegraphics[width=\textwidth]{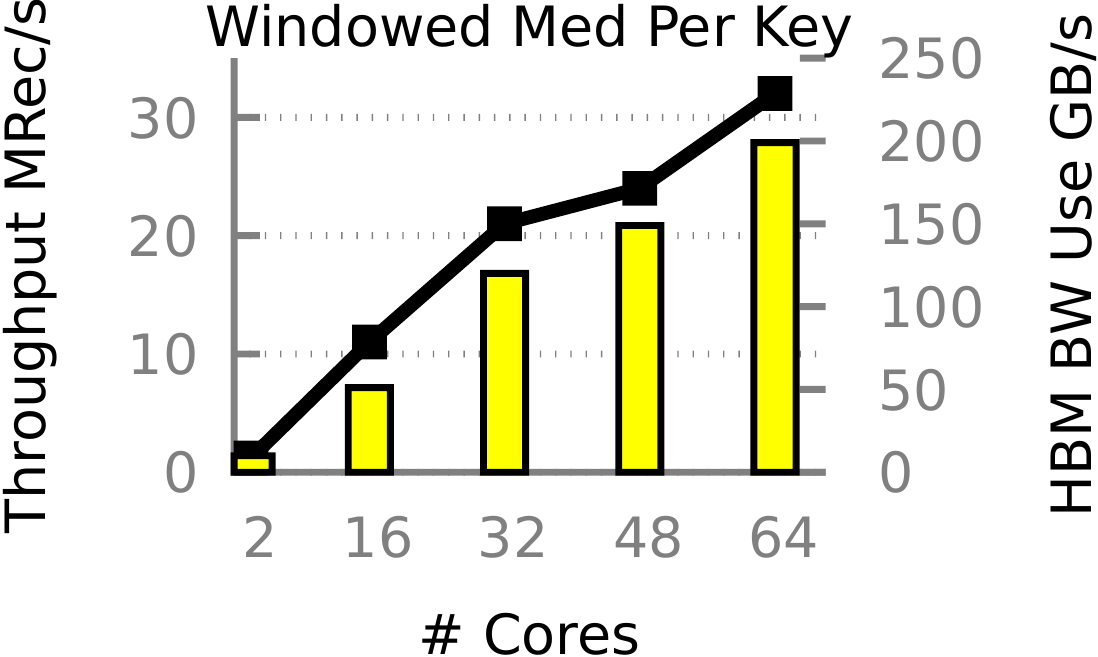}
%        \caption{wc \makalu{}}
        \label{fig:tiger}
    \end{subfigure}
    %~ %add desired spacing between images, e. g. ~, \quad, \qquad, \hfill etc. 
    \begin{subfigure}[b]{0.3\textwidth}
        \includegraphics[width=\textwidth]{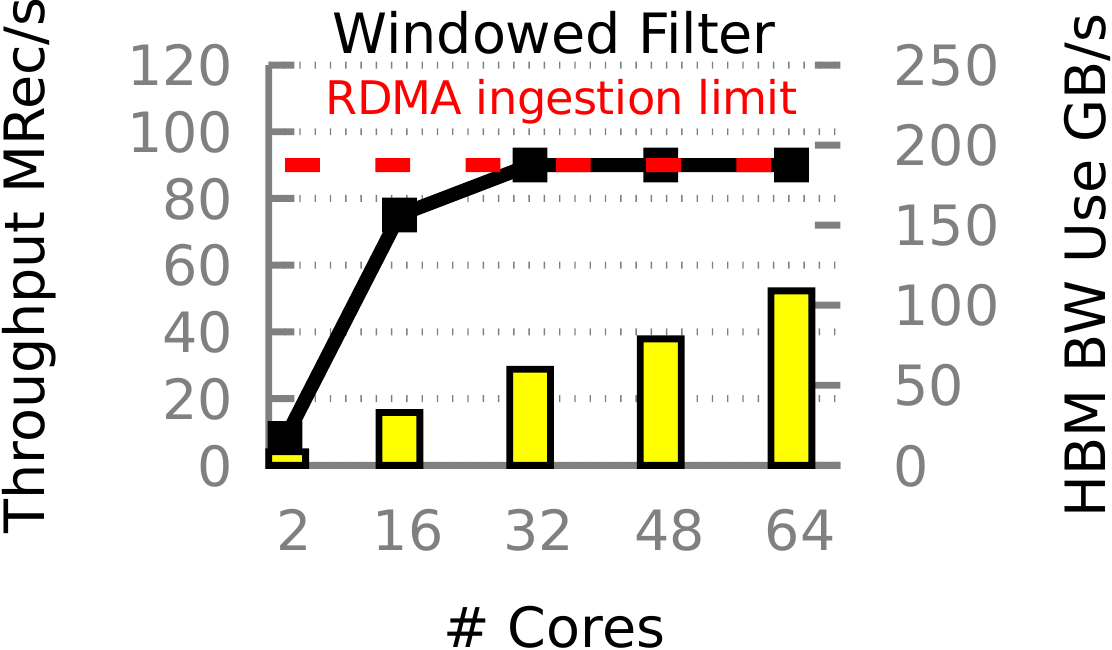}
%        \caption{wc \ktwo{}}
        \label{fig:mouse}
    \end{subfigure}
    %~
    \begin{subfigure}[b]{0.3\textwidth}
        \includegraphics[width=\textwidth]{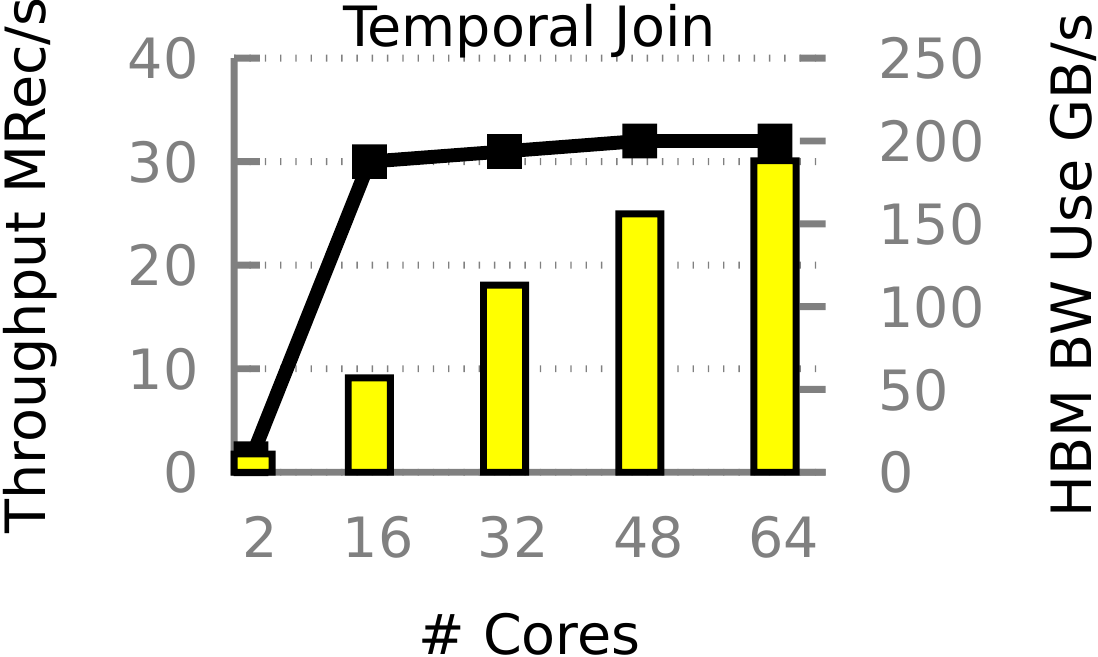}
%        \caption{wc \ktwo{}}
        \label{fig:mouse}
    \end{subfigure}    
    %~
    \begin{subfigure}[b]{0.3\textwidth}
        \includegraphics[width=\textwidth]{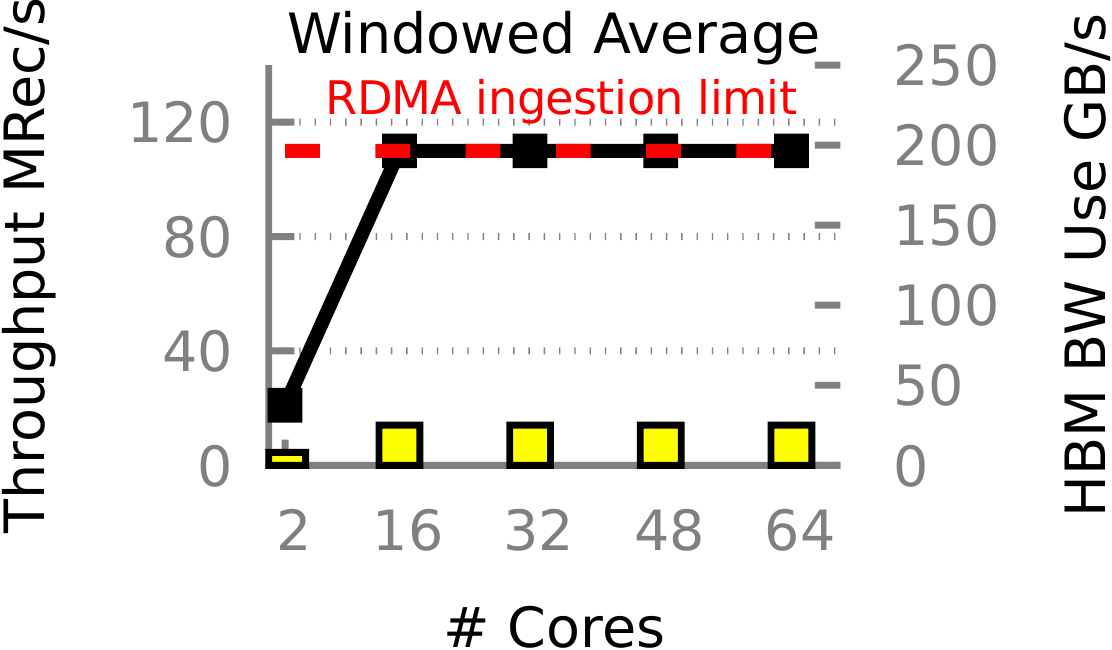}
%        \caption{wc \ktwo{}}
        \label{fig:mouse}
    \end{subfigure}    
    %~
    \begin{subfigure}[b]{0.3\textwidth}
        \includegraphics[width=\textwidth]{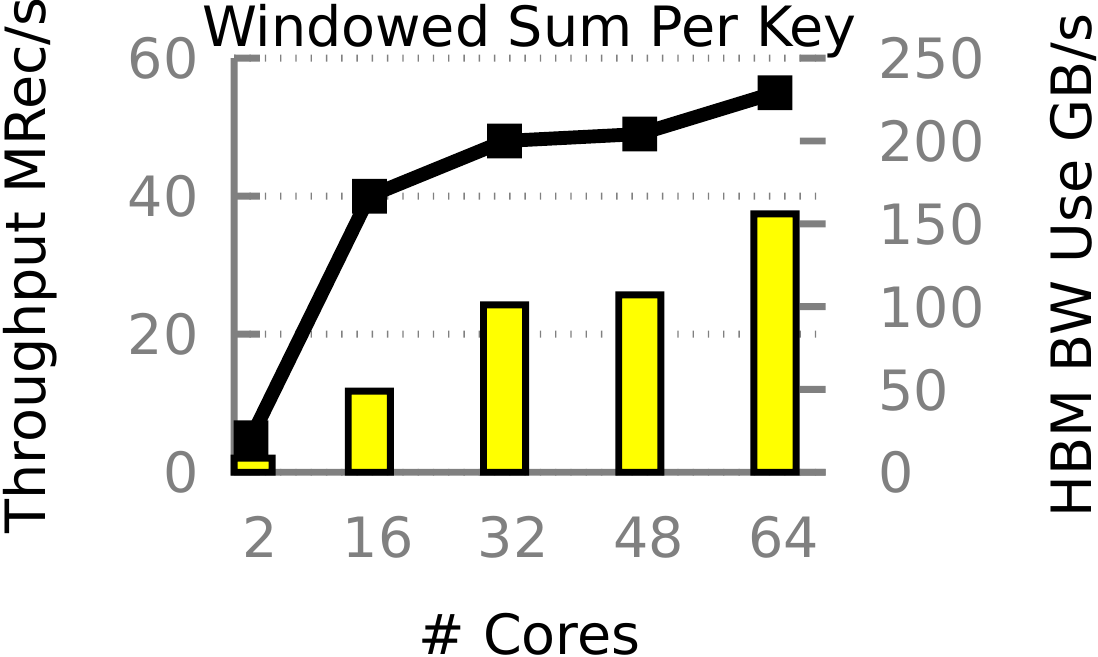}
%        \caption{wc \ktwo{}}
        \label{fig:mouse}
    \end{subfigure}    
    %~
    \begin{subfigure}[b]{0.3\textwidth}
        \includegraphics[width=\textwidth]{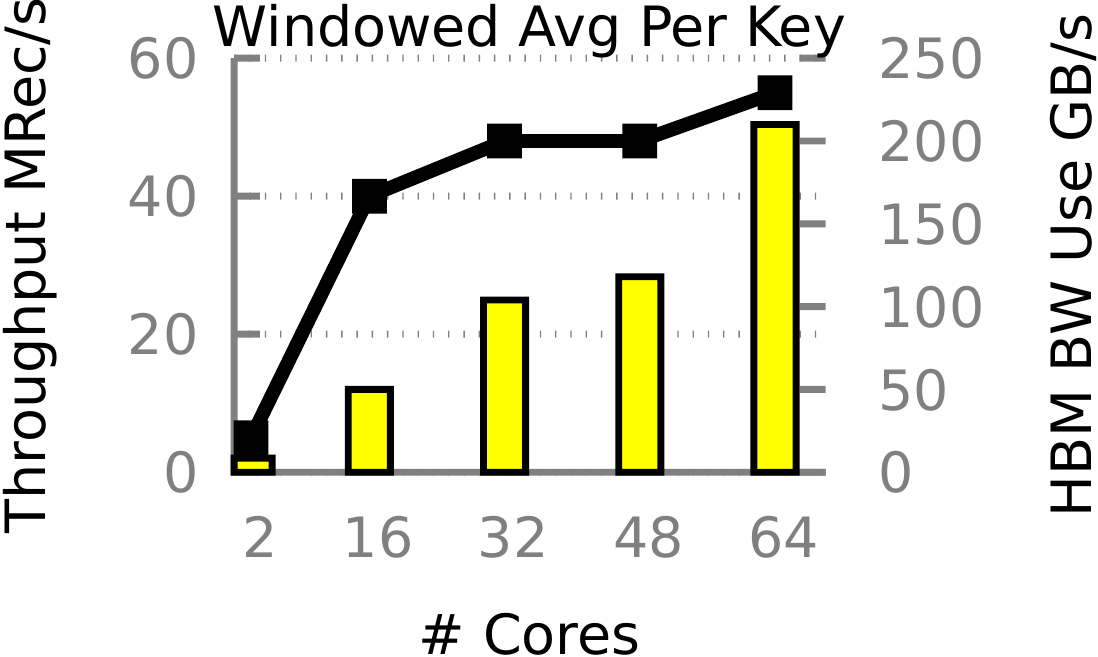}
    %        \caption{wc \ktwo{}}
        \label{fig:mouse}
        \end{subfigure}
	%~
    \begin{subfigure}[b]{0.3\textwidth}
        \includegraphics[width=\textwidth]{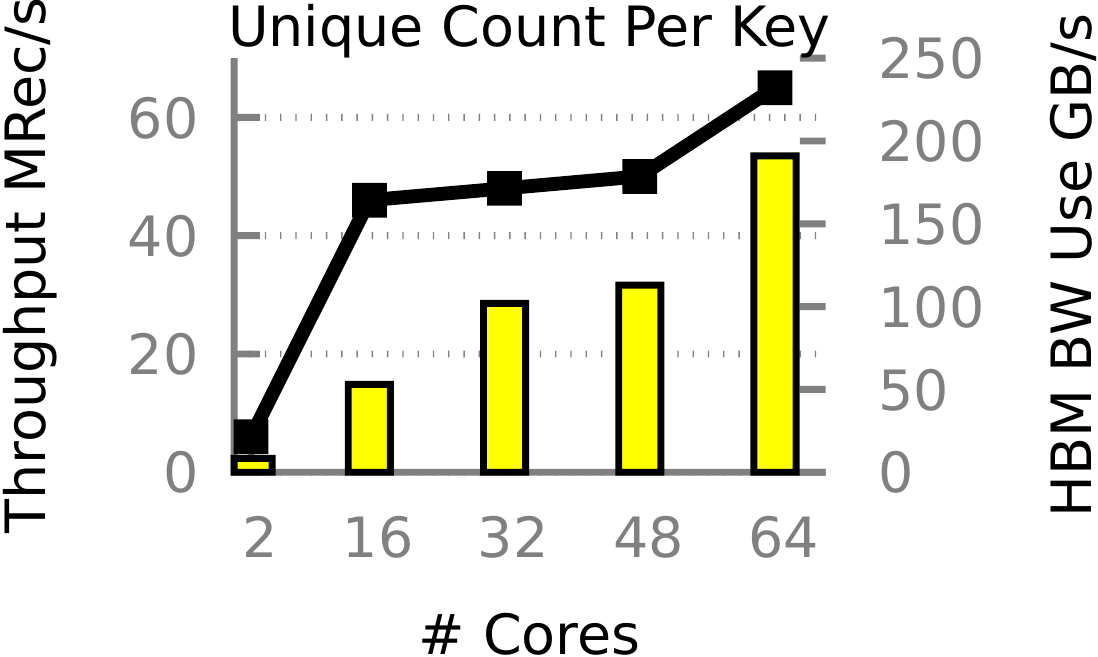}
    %        \caption{wc \ktwo{}}
        \label{fig:mouse}
        \end{subfigure}  
	%~
    \begin{subfigure}[b]{0.3\textwidth}
        \includegraphics[width=\textwidth]{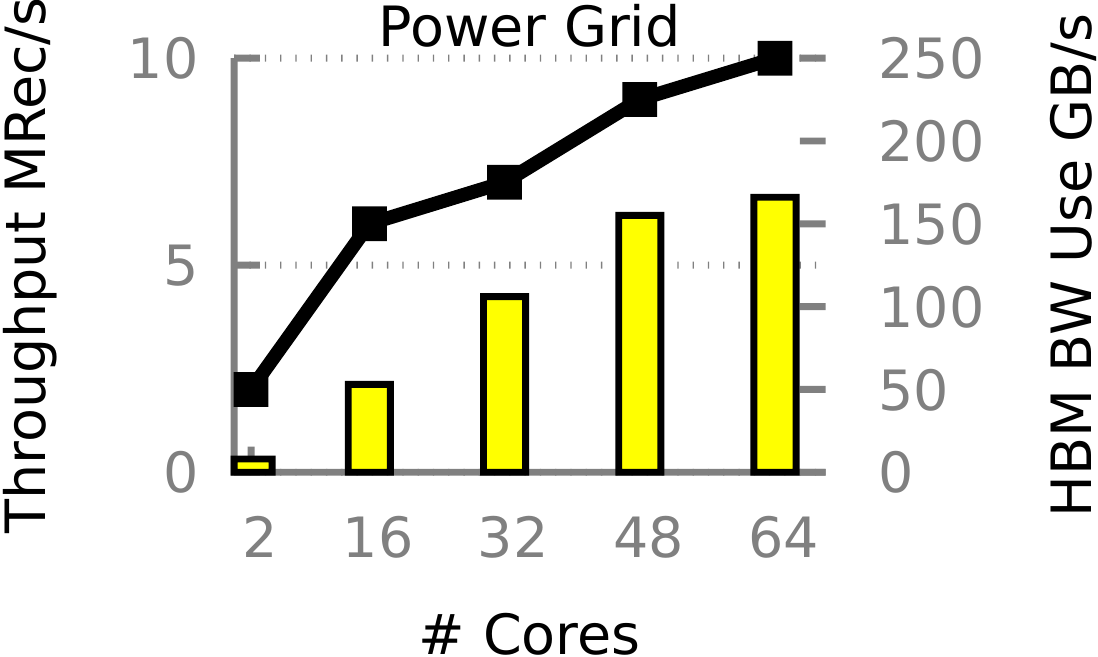}
    %        \caption{wc \ktwo{}}
        \label{fig:mouse}
        \end{subfigure}    
    \vspace{-10pt}		% use as needed

    \caption{\sys{}'s throughput (as lines, y-axis on left) and peak bandwidth utilization of HBM (as columns, y-axis on right) under 1-second target output delay. \sys{} shows good throughput and high memory bandwidth usage %\textbf{tput in GB/sec}
    }
    \label{fig:scalability}
    
    %\vspace{-5pt}		% use as needed
    
\end{figure*}

%\subsection{Throughput and Memory Bandwidth Utilization}
\subsection{Throughput and Bandwidth}
\label{sec:eval:tput}
%\subsection{Performance on 9 Benchmarks}
%As YCSB is simple, for which \sys{} easily saturates IO, 
We use nine benchmarks and experimental setup described in Section~\ref{sec:impl} to demonstrate that \sys{}: (1) supports simple and complex pipelines,
(2) well utilizes HBM bandwidth,
and (3) scales well for most pipelines.
%To explore the performance of \sys{} on a variety of stream pipelines, 

\paragraph{Throughput and scalability} 
Figure~\ref{fig:scalability} shows throughput on the left y-axis as a function of hardware parallelism (cores) on the x-axis. 
\sys{} delivers high throughput and processes between 10 to 110 M records/s while keeping output delay under the 1-second target delay.  Six benchmarks scale well with hardware parallelism and three benchmarks achieve their maximum throughput at 16 or 32 cores. Scalability diminishes over 16 cores in a few benchmarks because the engine saturates 
%the 40$\;$Gb/s Infiniband IO 
RDMA ingestion  (marked as red horizontal lines in the figures). 
 Most other benchmarks range between 10 and 60 M records/sec. 
The simple Windowed Average pipeline achieves 110 M records/sec (2.6 GB/s) with 16 participating cores. 
%saturating the 40Gb/s Infiniband IO we use. 
\sys{}'s good performance is due its effective use of HBM and its creation and management of parallelism.

%Figure~\ref{fig:scalability} shows \sys{} scales well with core count in most of the benchmarks. 

\paragraph{Memory bandwidth utilization}  
\sys{} generally utilizes HBM bandwidth well. 
When all 64 cores participate, most benchmarks consume 150--250 GB/sec, which is 40\%--70\% of the HBM  bandwidth limit. %\note{check} 
Furthermore,  the throughput of most benchmarks benefits from this bandwidth, which far exceeds the machine's DRAM peak bandwidth (80 GB/sec). 
Profiling shows that bandwidth is primarily consumed by Sort and Merge primitives for data grouping. 
A few benchmarks show modest memory bandwidth use, because their computations are simple and their pipeline are bound by the IO throughput of ingestion.

\subsection{Demonstration of Key Design Features}
\label{sec:eval:valid}

This section compares software and hardware \sys{} configurations, demonstrating their performance contributions.

\newcommand{\sysdram}{\sys{} DRAM}
\newcommand{\sysfr}{\sys{} Caching NoKPA}
\newcommand{\syscache}{\sys{} Caching}

\begin{figure}
    \centering
	    	\includegraphics[width=0.4\textwidth{}]{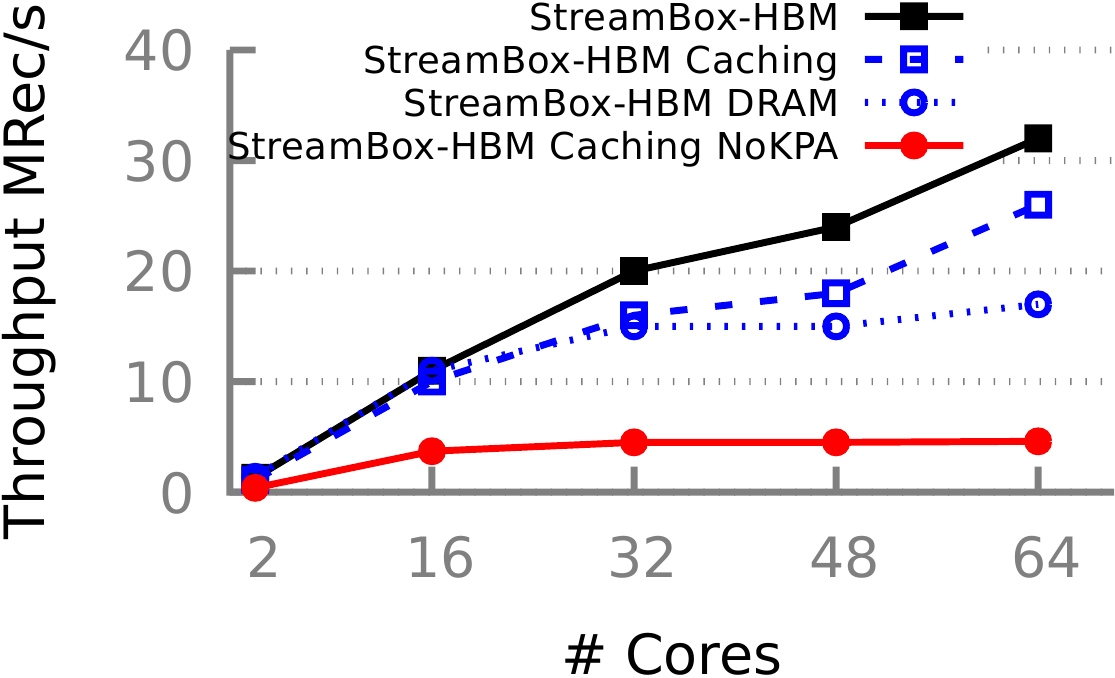}
	    %\caption{\wc{}}
	    %\label{fig:cmp_full_rec}
%    \begin{subfigure}[b]{0.2\textwidth}
%        \includegraphics[width=\textwidth]{figs/plot/cmp_flink/cmp_tput.pdf}
%        \caption{\wc{}}
%        \label{fig:gull}
%    \end{subfigure}
%    ~ %add desired spacing between images, e. g. ~, \quad, \qquad, \hfill etc. 
%      %(or a blank line to force the subfigure onto a new line)
%    \begin{subfigure}[b]{0.2\textwidth}
%        \includegraphics[width=\textwidth]{figs/plot/cmp_flink/cmp_bw.pdf}
%        \caption{\netmon{}}
%        \label{fig:tiger}
%    \end{subfigure}
    \caption{\sys{} outperforms alternative implementations, showing the efficacy of \hbma{} and its management of hybrid memory. Benchmark: TopK Per Key}\label{fig:cmp_full_rec}
   % \vspace{-8pt}		% use as needed
\end{figure}

\paragraph{HBM hardware benefits}
% \sys{} critically depends on its exploitation of HBM. 
To show HBM benefits versus other changes, we configure our system to use only DRAM  (\sysdram{}) and compare to \sys{} in
Figure~\ref{fig:cmp_full_rec}.  \sysdram{} reduces throughput by 47\% versus \sys{}.
Profiling reveals performance is capped due to saturated DRAM bandwidth. 

\paragraph{Efficacy of \hbma{}}
We demonstrate the extraction benefits of \hbma{} on HBM by modifying the engine to operate on full records.
Because HBM cannot hold all streaming data, we use cache mode, thus relying on the hardware to migrate the data between HBM and DRAM (\sysfr). 
This configuration still uses sequential-access computations, just not on extracted KPA records.
It is StreamBox~\cite{streambox} with sequential algorithms on hardware-managed hybrid memory.
Figure~\ref{fig:cmp_full_rec} shows  \sys{} outperforms \sysfr{} 
consistently on all core counts by up to 7$\times$. 
Without KPA and software management of HBM, scaling is limited to 32 cores.
The performance bottleneck is excessive data movement due to migration and grouping full records. 

% !TeX root = main.tex

% !TeX root = main.tex

%\begin{figure}
%    \centering
%    %\includegraphics[width=0.45\textwidth{}]{figs/plot/comparison/spark-beam}
%	    \includegraphics[width=0.45\textwidth{}]{figs/plot/mem_pressure/change_ingestion/change_ingestion.pdf}
%	    %\caption{\wc{}}
%	    %\label{fig:cmp_full_rec}
%%    \begin{subfigure}[b]{0.2\textwidth}
%%        \includegraphics[width=\textwidth]{figs/plot/cmp_flink/cmp_tput.pdf}
%%        \caption{\wc{}}
%%        \label{fig:gull}
%%    \end{subfigure}
%%    ~ %add desired spacing between images, e. g. ~, \quad, \qquad, \hfill etc. 
%%      %(or a blank line to force the subfigure onto a new line)
%%    \begin{subfigure}[b]{0.2\textwidth}
%%        \includegraphics[width=\textwidth]{figs/plot/cmp_flink/cmp_bw.pdf}
%%        \caption{\netmon{}}
%%        \label{fig:tiger}
%%    \end{subfigure}
%    \caption{\sys{} can adapt to different ingestion rates and balance the pressure of two types of memory to meet their constrains. }\label{fig:atativity_ingest}
%\end{figure}
%
%\begin{figure}
%    \centering
%    %\includegraphics[width=0.45\textwidth{}]{figs/plot/comparison/spark-beam}
%	    \includegraphics[width=0.45\textwidth{}]{figs/plot/mem_pressure/change_bundles_per_epoch//change_bundles_per_epoch.pdf}
%
%    \caption{\sys{} can adapt to different bundles per epoch. }\label{fig:adapt_bd_epoch}
%\end{figure}

\begin{figure}
    \centering
	    \includegraphics[width=0.45\textwidth{}]{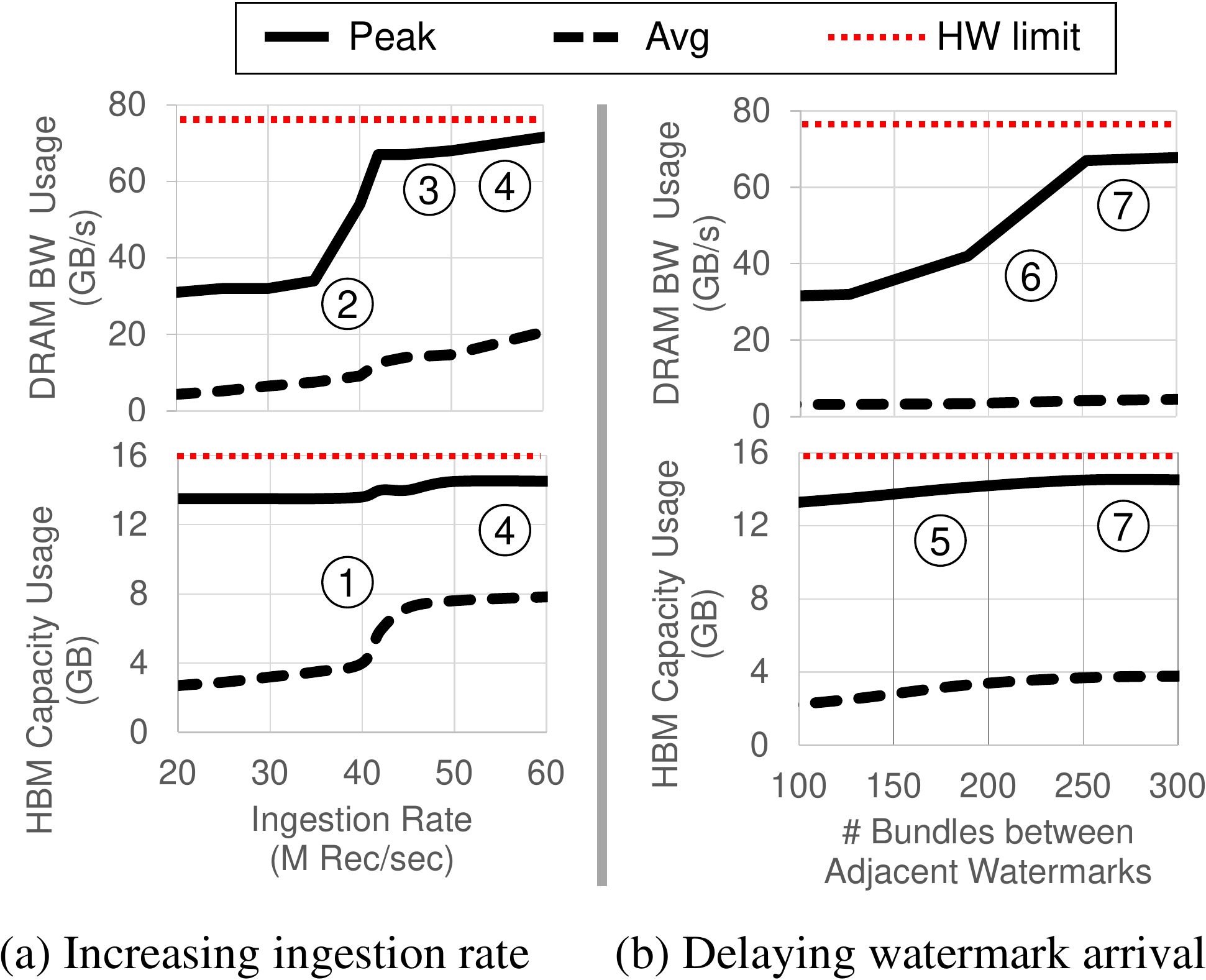}
    \caption{\sys{} dynamically balances its demands for limited memory resources under varying workloads. 
    Benchmark: TopK Per Key
    }\label{fig:adapt}
    %\vspace{-5pt}		% use as needed
\end{figure}

% Argument: KRP is more efficient and faster than full records. KRP requires fewer memory movement. 
%Methodology: implement two version of TopKperV pipeline. v1 uses our KRP mechanism, and v2 works on full records. We run v1 and v2 in cache mode, and compare their throughputs and memory bandwidth utilization. 
%Figure: x-axis shows the number of cores; y1-axis shows v1's throughput, y2-axis shows v2's throughput; y3-axis shows v1's memory bandwidth utilization, y4-axis shows v2's memory bandwidth utilization. 

%\paragraph{Engine-controlled data placement}
\paragraph{Explicit \hbma{} placement}
\sys{} fully controls \hbma{} placement and eschews transparent management by the OS or hardware. 
To show this benefit, we run \hbma{} by turning off \hbma{} placement and configuring HBM and DRAM in cache mode (\syscache{}). 
This configuration still enjoys the \hbma{} mechanisms, but relies on hardware caching to migrate \hbma{}s between DRAM and HBM.
Figure~\ref{fig:cmp_full_rec} shows \syscache{} drops throughput up to 23\% compared to \sys{}. 
The performance loss is due to excessive copying. All \hbma{}s must be first instantiated in DRAM before moving to HBM.
The hardware may move full records to HBM, paying a cost while having little performance return. 
For stream processing, software manges hybrid memories better than hardware.

\begin{comment}
\noindent 
{\bf Software is better than hardware to manage hybrid memory}

Argument: it is better to let the software to manage hybrid memory than hardware(cache mode). Reason 1: software (stream engine) itself have better knowledge than hardware about the memory usage pattern. Reason 2: hardware management is not efficient because all data has to be loaded from DRAM to HBM in cache mode, which is limited by DRAM bandwidth and introduces lots of migration overhead. While our software management never loads full records to HBM.  

Methodology: run TopKperV pipeline in two versions, which are v1 and v2. Both of them use our KRP mechanism. v1 runs in flat mode, and the engine manages the hybrid memory explicitly; v2 runs in cache mode, and hardware manages the hybrid memory. 

Figure: x-axis shows the number of cores; y-1 shows the v1's throughput, y-2 shows v2's throughput; y-3 shows v1's memory bandwidth utilization, y-4 shows v2's memory bandwidth utilization.
\end{comment}

\paragraph{Balancing memory demands}
%Note: here the througput of topK is higher than the throughput in Figure 8(win: 1000* 1000), because we are using smaller window here (500*1000)
To show how \sys{} balances hybrid memory demands dynamically, 
 we increase data ingress rates to increase memory usage. % note: we can't keep under the target delay
Figure~\ref{fig:adapt}a shows when we increase the ingestion rate, HBM capacity usage increases \circlew{1}. 
\sys{} kicks in to counterbalance the trend, allocating more \hbma{}s on DRAM \circlew{2}. 
Computation on the extra \hbma{}s on DRAM substantially increases DRAM bandwidth utilization. \sys{} controls the peak value at 70 GB/sec, close to the DRAM bandwidth limit without saturating it \circlew{3}. 
As ingestion rate increases, \sys{} keeps both resources highly utilized without exhausting them by adding back pressure to ingestion \circlew{4}.
Figure~\ref{fig:adapt}b shows when we delay ingestion watermarks, which extends \hbma{} lifespans in HBM, adding pressure on HBM capacity \circlew{5}. 
Observing the increased pressure, \sys{} allocates more \hbma{}s on DRAM, which increases DRAM bandwidth usage \circlew{6}. 
As pressure on both resources  increases, \sys{} keeps utilization of both high without exhausting them \circlew{7}.

\begin{figure}
    \centering
	\includegraphics[width=0.45\textwidth{}]{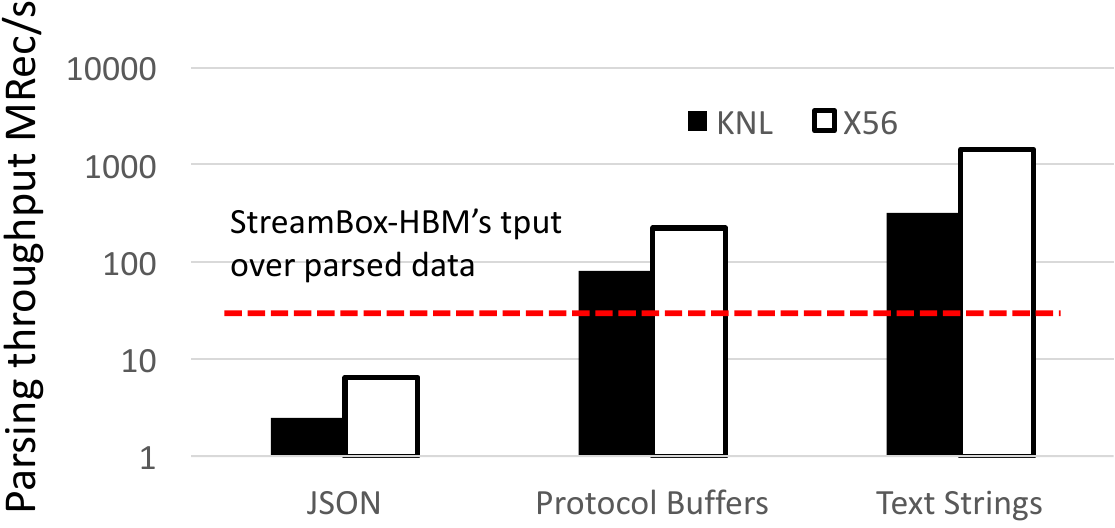}
    \caption{Parsing at the ingestion shows varying impacts on the system throughput. All cores on KNL and X56 are in use.  Parsers: RapidJSON~\cite{rapidjson}, Protocol Buffers (v3.6.0)~\cite{protoco-buffers}, and text strings to uint64~\cite{string_to_uint64}. 
    Benchmark: YSB
    }
    \label{fig:parsing-tput}
    %\vspace{-8pt}		% use as needed
\end{figure}

\begin{comment}
\begin{table}[t!]
	\centering
	\includegraphics[width=0.47\textwidth]{./figs/parsing-tput}
	\caption{Throughputs of parsing YSB records in MRec/s. XXX List pb and json libs info here. e.g. (C++, v3.6.0) XXX all cores in use XXX}
	%\vspace*{-15pt}
	%\vspace{-2pt}		% use as needed
	\label{tab:parsing-tput}
\end{table}
\end{comment}

\subsection{Impact of Data Parsing at Ingestion}
Our design and evaluation so far focus on a common situation where the engine ingests and processes numerical data~\cite{tersecades}. 
Yet, some streaming systems may ingest encoded data, parsing the data before processing. 
To examine how data parsing would impact \sys{}'s throughput, we construct microbenchmarks that parse the encoded input for the YSB benchmark. 
We tested three popular encoding formats:
JSON, Google's Protocol Buffers, and simple text strings.
We run these microbenchmarks on KNL and X56 (listed in Table~\ref{tab:plat}) %, and compare the parsing throughputs to \sys{}'s throughput on YSB.
to see if the parsing throughputs can keep up with \sys{}'s throughput on YSB.

As shown in Figure~\ref{fig:parsing-tput}, parsing at the ingestion shows varying impacts, depending on the ingested data format. 
While parsing simple text strings can be 29%1449/50=29 makalu
$\times$ as fast as \sys{} processing the parsed numerical data, 
parsing protocol buffers is 4.4%221/50=4.4 
$\times$ as fast, 
and parsing JSON is only 0.13%6.54/51 = 0.13 
$\times$ as fast. 
Our results also show that data parsing on X56 is 3-4$\times$ faster than KNL in general. 

Our results therefore have two implications towards fast stream processing when ingested data must be parsed first.
First, one shall consider avoiding ingested data formats (e.g. JSON) 
that favor  human-readability over efficient parsing.
%without efficient parsing. 
Data in such formats shall be transcoded near the data sources. 
Second, since KNL excels at processing numerical data but is disadvantaged in data parsing,
system administrators may team up Xeon and KNL machines as a hybrid cluster: the Xeon machines parse ingested data and the KNL machines run \sys{} to execute the subsequent streaming pipeline.

\section{Related Work}
\label{sec:related}

%\note{cite own work -- profdp and streambox. edgeflow?? cite x-stream}

% --- about TerseCades: ---
%TerseCades~\cite{tersecades} supports query execution on compressed stream data to exploit existing data redundancy, but only applies it to certain types of operators that are applied to numerical data. TerseCades optimizations are orthogonal to that of KPA and both techniques can be used in one holistic design.

% TerseCades~\cite{tersecades} explores stream data compression and direct execution on the compressed data for a certain class of operators. It is orthogonal and can potentially benefit HyStream.

\paragraph{Stream analytics engines}
%\note{cite a few classic engines?}
Much prior work improves stream analytics performance on a single node.
%Zhang et al.~\cite{zhang17icde} improve inefficiency originated from parallel execution model and NUMA-awareness of commodity engines when executing on a multicore server.
StreamBox coordinates task and data parallelism with a novel out-of-order bundle processing approach, achieving high throughput and low latency on multicores~\cite{streambox}. 
%TerseCades~\cite{tersecades} executes queries on compressed stream data to exploit data redundancy.
SABER accelerates streaming operators using multicore CPU and GPU~\cite{saber}.
Other work uses FPGA for stream processing~\cite{hagiescu2009computing}.
No prior work, however, optimizes stream analytics for hybrid memories. %as \sys{} does.
\sys{} complements prior work that addresses diverse needs in distributed stream processing~\cite{google-dataflow,streamscope, naiad,timestream,toshniwal2014storm,sparkstreaming}. They address issues such as fault tolerance~\cite{streamscope,timestream,sparkstreaming}, programming models~\cite{naiad}, and adaptability~\cite{gloss,drizzle}.
As high throughput is fundamental to distributed processing,
\sys{} can potentially benefit those systems regardless of their query distribution methods among nodes.

%\sys{} is related to investigation on single-node stream analytics performance.
%Zhang et al.~\cite{zhang17icde} show substantial inefficiency in executing commodity engines on modern multicores.
%SABER~\cite{saber} accelerates streaming operators by multicore CPU and GPU.
%StreamBox~\cite{streambox} processes out-of-order stream epochs for higher parallelism.
%Trill~\cite{trill} stores and processes records in columnar format.
%\sys{} also complements work that addresses challenges in scaling out stream analytics.
%They adderess challenges including fault tolerance~\cite{sparkstreaming,timestream,streamscope}, programming models~\cite{naiad}, and adaptability~\cite{drizzle,gloss}.
%None prior work, however, optimizes stream analytics for hybrid memories as \sys{} does.

\paragraph{Managing keys and values}
\hbma{} is inspired by key/value separation~\cite{alphasort}.
Many relational databases store records in columnar format~\cite{monetdb,sqlserver-column,db2-column,c-store} or use an in-memory index~\cite{lehman1986query} to improve data locality and speed up query execution.
For instance, Trill applies columnar format to bundles to efficient process only accessed columns, but extracts all of them at once~\cite{trill}.
% Similarly, \hbma{} improves spatial locality in stream processing.
%However, while prior work optimizes fast aggregation over data (e.g., by encoding them), \sys{} optimizes grouping within \hbma{}s with different techniques such as vectorized merge-sort.
Most prior work targets batch processing and therefore extracts columns ahead of time. 
%Furthermore, \sys{} creates \hbma{}s dynamically and selectively -- only for columns used to group keys.
By contrast, \sys{} creates \hbma{}s dynamically and selectively -- only for columns used to group keys. It swaps keys as needed, maintaining only one key from a record in HBM at time to minimize the HBM footprint. Furthermore, \sys{} dynamically places \hbma{}s in HBM and DRAM based on resource usage.

\paragraph{Data processing for high memory bandwidth}
X-Stream accelerates graph processing with sequential access~\cite{xstream}.
Recent work optimized quick sort~\cite{bramas2017fast}, hash joins~\cite{cheng17cikm}, scientific workloads~\cite{li17sc,7965110}, and machine learning~\cite{you17sc} for KNL's HBM, but not streaming analytics.
Beyond KNL, Mondrian~\cite{mondrian} uses hardware support for analytics on high memory bandwidth in near-memory processing.
Together, these results highlight the significance of sequential access and vectorized algorithms, affirming \sys{}'s design.

\paragraph{Managing hybrid memory or storage}
Many generic systems manage hybrid memory and storage.
X-mem automatically places application data based on application execution patterns~\cite{x-mem}. % and their memory access.
%They monitor the application behaviors and memory access patterns and therefore decide the placement of app data structures (as whole).
%Shoal~\cite{shoal} proposes an array abstraction and transparently move/replicate data among NUMA domains (\note{may not be directly related}).
Thermostat transparently migrates memory pages between DRAM and NVM while considering page granularity and performance~\cite{thermostat}.
CoMerge makes concurrent applications share heterogeneous memory tiers based on their potential benefit from fast memory tiers~\cite{comerge}.
Tools such as ProfDP measure performance sensitivity of data to memory location and accordingly assist programmers in data placement~\cite{profdp}.
Unlike these systems that seek to make hybrid memories transparent to applications, 
\sys{} constructs \hbma{}s specifically for HBM  and fully controls data placement for stream analytics workloads.
%While \sys{} does not have programmers involved, it has a full control over data placement similar to X-mem, Thermostat, and CoMerge.
%The key distinction is that \sys{} targets on \emph{plat} hybrid memory, where two memory types show different advantages between latency and bandwidth, as described in \sect{bkgnd}.
%\paragraph{Software for hybrid memory or storage}
%Many systems seek to manage hybrid memory/storage transparently while keeping applications oblivious.
%X-mem~\cite{x-mem} automatically places application data based on its observation of application execution patterns and memory access.
%%They monitor the application behaviors and memory access patterns and therefore decide the placement of app data structures (as whole).
%%Shoal~\cite{shoal} proposes an array abstraction and transparently move/replicate data among NUMA domains (\note{may not be directly related}).
%Thermostat~\cite{thermostat} transparently migrates app memory pages between DRAM and NVM while considering page granularity.
%CoMerge~\cite{comerge} makes concurrent apps share heterogeneous memory tiers based on potential benefit from fast memory.
%Tools such as ProfDP~\cite{profdp} measure memory objects' sensitivity to memory types and accordingly assist programmers in deciding data placement.
%Unlike them, \sys{} designs \hbma{}s specifically for HBM and fully controls data placement.
Several projects construct analytics and storage software for hybrid memory/storage~\cite{wisckey,hikv}.
Most of them target DRAM with NVM or SSD with HDD, where high-bandwidth memory/storage delivers lower latency as well.
Because HBM lacks a latency advantage, borrowing from these designs is not appropriate.
% It instead provides \hbma{} constructs and primitives to allow parallel execution on limited HBM capacity.

%Several projects construct analytics or storage software to exploit memory/storage heterogeneity~\cite{wisckey,hikv}.
%Most of them target DRAM with NVM or SSD with HDD while none targets HBM/DRAM, to the best of our knowledge.
%As described in \sect{bkgnd}, HBM, unlike such hybrid memory/storage, offers no latency advatange.
%This makes \sys{} unable to borrow existing designs but instead must provide its \hbma{} construct and primitives.

%\note{discuss: why can't NVM/DRAM be applied?}
%For instance, HiKV~\cite{hikv} optimizes kv-store for DRAM/NVM. It places B+ tree index in DRAM (fast write, volatile) and hash index in NVM (slow write, for eventual persistence).

%\note{discuss: App-aware approach}
%Often annotation, sensitivity, etc. -- should we criticize these here?

%\paragraph{Limitation} Current implementation of \sys{} only supports numerical data, which is very common in big data analytics. We will support more data types in future work. 
% !TeX root = main.tex

\section{Conclusions}
We present the first stream analytics engine that optimizes performance for hybrid HBM-DRAM memories.
%The design is challenging because HBM capacity is limited and because it only accelerates sequential-access, high parallelism workloads. 
Our design addresses the limited capacity of HBM and HBM's need for sequential-access and high parallelism. 
Our system design includes (i)  novel dynamic key / record pointer extraction into \hbma{}s that minimizes the use of precious HBM capacity,  
%it performs data grouping, the most demanding workloads in stream analytics, as merge-sort in HBM; 
(ii) sequential grouping algorithms on  \hbma{}s to balance limited capacity while  exploiting high bandwidth; and (iii) a  runtime that manages	 parallelism and \hbma{} placement in hybrid memories. 
%it reconstructs grouping computations 
%the most demanding workloads in stream analytics, 
%into sequential access by using sort-merge to benefit the high bandwidth of HBM; 
%it compresses stream records to be \hbma{} in HBM; 
%it compresses streaming records to be \hbma{} to use the high bandwidth of HBM while respects its capacity limit;
%it dynamically manages parallelism and memory, balancing the needs for two major resource limits: HBM's capacity and DRAM's bandwidth. 
%it dynamically manages parallelism and \hbma{} placement, balancing the needs for two major resource limits of HBM's capacity and DRAM's bandwidth. 
\sys{} achieves 110$\;$M records/second on a 64 core KNL machine. 
It outperforms engines without KPA and with sequential-access algorithms by 7$\times$ 
%in throughput and stream engines with random access algorithms by 18$\times$ per core throughput.
and engines with random-access algorithms by an order of magnitude. We find that for stream analytics, software better manages hybrid memories than hardware.

% We believe \sys{} is the first stream engine for hybrid memories.
% !TeX root = main.tex

%\section*{Acknowledgments}
\begin{acks}
For this project: the authors affiliated with Purdue ECE were supported in part by NSF Award 1718702, NSF Award 1619075, and
a Google Faculty Award.
%\note{Check with co-authors for what to add}
The authors thank the anonymous reviewers for their feedback. 
The authors thank Prof. Yiying Zhang for her technical suggestion and support. 
Heejin Park and Shuang Zhai contributed discussion on an early version of the paper.
\end{acks}
%Thank Yiying, Heejin, and Shuang.

\balance
%\footnotesize 
\bibliographystyle{acm}
%\bibliography{../common/bibliography}}
%\bibliographystyle{bib/abbrv-minimal}
\bibliography{bib/abr-long,bib/xzl,bib/iot,bib/misc,bib/book,bib/hongyu,bib/datacentric}
%\theendnotes

\end{document}